\documentclass[useAMS,usenatbib]{mn2e}

\usepackage{graphicx}
\usepackage{txfonts}
\def\muHz{\,\mu {\rm Hz}}
\def\note #1]{{\bf #1]}}
\def\ie{{\rm i.e.}}
\def\nablarad{\nabla_{\rm rad}}
\def\nablaad{\nabla_{\rm ad}}

\def\nablacsq{\nabla_{c^2}}

\def\taud{\tau_{\rm d}}
\def\btaud{\bar{\tau}_{\rm d}}
\def\bgamd{\bar{\gamma}_{\rm d}}
\def\deld{\Delta_{\rm d}}
\def\A25{A_{2.5}}

\def\CFovs{{\cal F}_{\rm ovs}}
\def\rt{r_{\rm t}}
\def\rf{r_{\rm f}}
\def\rd{r_{\rm d}}
\def\rcz{r_{\rm cz}}
\def\zetaf{\zeta_{\rm f}}
\def\dd{{\rm d}}

\def\dis{\displaystyle}
\def\diff{{\rm d}}
\def\CH{{\cal H}}
\def\CF{{\cal F}}
\def\MS{S}
\def\MSpr{{${\rm S}^{'}$}}
\def\OAone{A1}
\def\OAtwo{A2}
\def\OBone{B1}
\def\OBtwo{B2}
\def\OBthree{B3}
\def\OCone{C1}
\def\OCtwo{C2}
\def\OCthree{C3}
\def\OD{OD}
\def\OP{OP}
\def\rbf{\rm}

\def\gwig{{\leavevmode\kern0.3em\raise.3ex\hbox{$>$}
\kern-0.8em\lower.7ex \hbox{$\sim$}\kern0.3em}}
\def\lesssim{{\leavevmode\kern0.3em\raise.3ex\hbox{$<$}
\kern-0.8em\lower.7ex \hbox{$\sim$}\kern0.3em}}

\def\apj{ApJ}%
\def\apjl{ApJ}%
\def\aap{A\&A}%
\def\mnras{MNRAS}%
\def\apss{ApSS}%

\title[Overshoot in the Sun]
{A more realistic representation of overshoot at the base of the solar convective envelope as seen by helioseismology}
\author[J. Christensen-Dalsgaard et al.]
{J.~Christensen-Dalsgaard$^{1,2}$\thanks{
E-mail:jcd@phys.au.dk; mario.monteiro@astro.up.pt;\hfill\break
rempel@ucar.edu; mjt@ucar.edu},
M.J.P.F.G.~Monteiro$^{3,4}$,
M.~Rempel$^{2}$, M.J.~Thompson$^{2,5}$\\
$^{1}$Danish AsteroSeismology Centre, and Department of Physics and Astronomy,
   Aarhus University, Ny Munkegade, 
   DK-8000 Aarhus C, Denmark\\
$^{2}$High Altitude Observatory, NCAR\thanks{The National Center
   for Atmospheric Research is operated by the University
   Corporation for Atmospheric Research under sponsorship of
   the National Science Foundation}, Boulder CO~80307-3000, USA\\
$^{3}$Centro de Astrof\'{\i}sica, Universidade do Porto,
   Rua das Estrelas, 4150-762 Porto, Portugal\\
$^{4}$Departamento de F\'{\i}sica e Astronomia,
   Faculdade de Ci\^encias, Universidade do Porto, Portugal\\
$^{5}$School of Mathematics \& Statistics, University of Sheffield,
   Hounsfield Road, Sheffield S3~7RH, United Kingdom}
\begin{document}

\date{Accepted ; Received ; in original form 2009 September 11}

\pagerange{\pageref{firstpage}--\pageref{lastpage}} \pubyear{2010}

\maketitle

\label{firstpage}

\begin{abstract}
The stratification near the base of the Sun's convective envelope is governed by processes of convective overshooting and element diffusion, and the region is widely believed to play a key role in the solar dynamo.
The stratification in that region gives rise to a characteristic signal in the frequencies of solar p modes, which has been used to determine the depth of the solar convection zone and to investigate the extent of convective overshoot. 
Previous helioseismic investigations have shown that the Sun's spherically symmetric stratification in this region is smoother than that in a standard solar model without overshooting, and have ruled out simple models incorporating overshooting, which extend the region of adiabatic stratification and have a more-or-less abrupt transition to subadiabatic stratification at the edge of the overshoot region. 
In this paper we consider physically motivated models which have a smooth transition in stratification bridging the region from the lower convection zone to the radiative interior beneath. We find that such a model is in better agreement with the helioseismic data than a standard solar model.
\end{abstract}


\section{Introduction}

\label{sec:intro}

An understanding of the overshoot region at the bottom of the Sun's convective envelope is important for a number of reasons.
The overshoot region approximately coincides with the solar tachocline, a region of rotational shear
which is generally believed to play a key role in the solar dynamo:
overshooting is likely to be important for helping to store the magnetic flux
below the convection zone during the solar cycle.
Bulk motion in the overshoot region also
affects the thermal stratification and it may contribute to significant
mixing of chemical elements, for example transporting fragile elements such
as lithium to hotter regions where they are destroyed more easily
than in the convection zone. More generally, convective overshoot in stars
(particularly those with convective cores) is likely to be an important and
as yet imperfectly understood process affecting age estimates of stars, and
so improved constraints on theories of overshooting obtained from a study of
how overshooting works in the solar case can be important for understanding
stars more widely.

Helioseismology provides a means of probing directly the conditions inside
the Sun, because the frequencies of resonant modes set up by acoustic waves
propagating in the solar interior depend in particular on the adiabatic sound
speed $c$ which is given by
\begin{equation}
c^2\ =\
{{\Gamma_1 p}\over{\rho}}\ \simeq\
{{\Gamma_1 k_{\rm B} T}\over{m_{\rm u}\mu}}\;.
\label{eq:apprcsq}
\end{equation}
Here $\Gamma_1$ 
is the logarithmic derivative of pressure $p$ with respect to
density $\rho$ at constant specific entropy, $T$ is temperature, $\mu$ is the
mean molecular weight, $k_{\rm B}$ is
Boltzmann's constant and $m_{\rm u}$ is the atomic mass unit.
Hence the sound-speed gradient with respect to depth
depends on the temperature gradient, which
itself depends on the mechanism by which heat is transported. The transition
between fully radiative heat transport beneath the convection zone
and convective heat transport within the convection zone is manifest
in the temperature gradient and hence too in the sound-speed gradient.
If the transition in sound-speed gradient takes place
over a distance that is small compared with the vertical
wavelength of the acoustic waves near the base of the convection zone, then
the transition appears to the waves to be more-or-less sharp and this gives
rise to an oscillatory signal in the mode frequencies $\omega$:
the form of the signal
gives information about the location and nature of this 
``acoustic glitch'' \citep{Gough2002a}.

\cite{Monteiro:etal:1994} represented the effect of the base of the convection
zone in terms of an additive contribution $\delta\omega_{\rm p}$ 
to the frequencies,
relative to those of a corresponding model in which the transition had been
smoothed out.
For low-degree modes (see \citealt{Monteiro:etal:2000}) a break in the
first derivative of the sound speed gives rise to a contribution of the form
\begin{equation}
\delta\omega_{\rm p} \ =\ A(\omega)\; \cos\left(
2\omega\bar\tau_{\rm d}+2\phi_0\right)\;, 
\end{equation}
while a break in the second derivative gives rise to a similar signal but 
with a sine term instead of a cosine, and with a different frequency dependence of $A(\omega)$.
Here $\bar\tau_{\rm d}$ is essentially 
the value of $\tau$ at the location of the acoustic glitch, where 
\begin{equation}
\label{eq:tau}
\tau = \int_r^R {\dd r' \over c} 
\end{equation}
is acoustic depth beneath the surface,
$r$ being the corresponding distance to the centre and $R$ the surface radius
of the Sun.
Also, $\phi_0$ is a phase introduced by the reflection of the mode at the turning points, depending in particular on
the near-surface structure. The function $A(\omega)$ is an amplitude which
depends on the sharpness and nature of the convection-zone base: the smoother
the transition, the smaller in general will be the amplitude.
However, if moderate-degree data are used, as in the case of Sun where we have accurate data for modes whose degree is above 3, the above expression needs to include additional terms, both in the amplitude (\citealt{Monteiro:etal:1994}) and in the argument of the signal (\citealt{ChristensenDalsgaard:etal:1995}), to account for the first-order effect of the mode degree on the signal.

Overshoot at the base of the solar convection has traditionally been modelled
using non-local mixing-length theory \citep[e.g.,][]{Zahn1991}.
Such models mostly predict an overshoot region that is nearly
adiabatically stratified;
in terms of the logarithmic temperature gradient 
$\nabla = \diff \ln T / \diff \ln p$ one finds that
$\delta\equiv\nabla-\nabla_{\rm ad}\sim-10^{-6}$, 
where $\nablaad = (\partial \ln T / \partial \ln p)_s$
is the adiabatic temperature gradient, $s$ being specific entropy.
The depth of overshoot region is
typically between $0.2H_p$ and $0.4H_p$,
where $H_p$ is the pressure scale height at the base
of the convection zone,
with a very steep transition towards the radiative temperature gradient.
The results have seemed rather robust, since the above behaviour is found in
models incorporating quite different large-scale flow structures: for example
\citet{Ballegooijen:1982a} assumed
overturning convective rolls whilst \citet{Schmitt:etal:1984}
explicitly modelled downward plumes in the overshoot region.
We note, however, that other treatments of the overshoot region have
suggested a much smoother transition to the radiative gradient
\citep[e.g.,][]{Xiong2001, Deng2008, Baturi2010}.

The rather abrupt transition in the temperature gradient
predicted by the non-local mixing-length models has been
parameterized and incorporated into solar models 
\citep{Basu:etal:1994, Basu1994, Monteiro:etal:1994,
Roxburgh:Vorontsov:1994, ChristensenDalsgaard:etal:1995}
in order to compare the predicted
acoustic-mode frequencies with those observed on the Sun.
Models with and without overshooting all have an oscillatory signal in 
frequencies coming from the base of the adiabatically stratified region, 
but the amplitude of the signal in the frequencies in the
overshoot models is greater than in models without overshooting.

When the observed and model frequencies are compared,
it is found that the amplitude of $\delta\omega_{\rm p}$ in the Sun
is comparable with or smaller than that in models without overshooting,
implying that the amount of overshooting of the kind predicted by these
mixing-length models is very small.
\cite{Monteiro:etal:1994} and \cite{ChristensenDalsgaard:etal:1995}
concluded that the extent of any such overshoot at the base of the convection
zone was less than one tenth of a pressure scale-height. A similar limit was
found by \cite{Basu:etal:1994}, while \cite{Roxburgh:Vorontsov:1994}
obtained a somewhat weaker limit.
\citet{Basu1994} noted that the composition gradient caused
by the inclusion of helium settling produced a sharper transition
in the sound-speed gradient, even in models without overshoot,
and hence a larger oscillatory signal.
   From these analyses it would appear that the transition in sound-speed gradient at the base of the solar convection zone is
if anything smoother than in the non-overshoot models. A caveat is that what we
purport to measure in the above studies
is the spherically symmetric component of the structure: departures from
sphericity, such as a
latitudinal dependence to the shape of the
base of the convection zone, could
make the transition appear smoother than it is locally. The helioseismic
evidence, however, is that the location of the base of the convection zone
is independent of latitude \citep{Monteiro:Thompson:1998,Basu:Antia:2001}.
Changes of the base of the convection zone on time scales shorter than the 
observation interval could also have a similar effect by introducing a 
time-averaging effect on the mode frequencies that would 
mimic a smoother transition.
The importance of this
effect
is difficult to estimate, however,
and depends strongly on the 3D nature of convection at the base of the 
envelope.

Overshoot has been addressed in the last two decades by
a variety of 2D and 3D numerical simulations \citep{Roxburgh:Simmons:1993,
Hurlburt:etal:1994,Singh:etal:1995,Singh:etal:1998,
Saikia:etal:2000,Brummell:etal:2002,Rogers:Glatzmaier:2005:ov,
Rogers:Glatzmaier:2005:grav}.
Whilst the non-local mixing-length models have clearly predicted an adiabatic
overshoot region of a sizeable fraction of a pressure scale height and a
rather sharp transition to the radiative zone beneath, the numerical
simulations show a greater variety of possible behaviours.
The work by \citet{Brummell:etal:2002} is
currently one of the best resolved and most turbulent investigations: it
shows
strongly subadiabatic overshoot with very smooth transition towards the
radiative temperature gradient.
Most of the earlier 2D and 3D simulation were
more in the laminar regime and found, depending on their parameters (mainly
the stiffness of the subadiabatic layer), both nearly adiabatic overshoot and
extended subadiabatic overshoot.

\citet{Rempel:2004} tried to
understand the cause of the discrepancies in different overshoot treatments,
which have been attributed either to over-simplifications in the mixing-length
approach and related models or to the fact that numerical simulations are not
in the correct parameter range. Using  a semi-analytic plume model
\citet{Rempel:2004} showed that the main differences result from
different values of the energy flux used in these models (more exactly the
energy flux divided by the filling factor of downflows, expressed by a
dimensionless number $\Phi\equiv
F/(f\,p\sqrt{p/\varrho})$, with the energy flux $F$,
the downflow filling factor $f$, pressure $p$ and density $\varrho$ at the base
of the convection zone).
The value of $\Phi$ ranges from $10^{-10}$ (mixing-length models)
to up to $10^{-2}$ (numerical simulations). The influence of the energy flux
(lower energy flux leading to more adiabatic overshoot) was already indicated
in the work of \citet{Brummell:etal:2002} and has been confirmed by
\citet{Kapyla2007}
by varying the energy flux over two orders of
magnitude (with the caveat that
lowering the energy flux in a numerical simulations typically decreases also
the degree of turbulence).
The main conclusion of the work of
\citet{Rempel:2004} is that non-local mixing-length models and current
numerical simulations mark two extremes in terms of the parameter $\Phi$ and
the Sun might have a solution somewhere in between, however most likely 
closer to the mixing-length regime unless the downflow filling factor at the
base of the convection zone is tiny. This leads to the possibility that solar
overshoot has a stratification close to adiabatic, but with a transition
towards the radiative gradient that is smoother than found in
non-local mixing-length models.

Accordingly, we have created a family of semi-analytic models
that display a range
of possible characteristics for the convection zone, which may give better
agreement with the helioseismic observations than the old non-local
mixing-length models. Moreover, helioseismology may determine which model
provides the best fit to the actual stratification of the Sun;
this may, we hope,
teach us something about the physics of convection and convective
overshooting in stellar interiors. 

The goal of the present paper is to discuss these models and compare
their seismic properties with those observed.
Section~2 presents the characteristics of the semi-analytic overshoot
model and its parameterization in our stellar structure calculations.
In Section~3 we describe the seismological method we use to analyse the 
oscillation frequencies to establish the characteristics of the 
transition in structure near the base of the convective envelope,
and we apply it to the analysis of solar data in Section~4.
Section~5 describes the solar models used in this study and presents
the analysis of the frequencies computed from those models. 
In Section~6 we synthesize the observational and model results and 
discuss our inferences, and we present our conclusions in Section~7.

\section[]{A simple parameterization for overshoot}
\label{sec:param}

{\rbf As summarized in the introduction a variety of overshoot models including
simplified plume models \citep{Rempel:2004}, non-local convection models
\citep[e.g.,][]{Xiong2001, Deng2008}, and 3D simulations 
\citep[e.g.,][]{Brummell:etal:2002} predict overshoot profiles that are
substantially smoother than those obtained through non-local mixing-length
models. While the physical reasons for the smoother transitions differ among 
these models, the resulting profiles show a large degree of similarity. 
For the purpose of the helioseismic investigation in this paper it is not essential to use exactly 
one of these models. Rather what is important is to realize that overshoot profiles smoother than 
those obtained by standard mixing-length theory are possible.
Using the models of \cite{Rempel:2004} as guidance 
we find the following properties to be rather general for all profiles:
\begin{enumerate}
   \item The overshoot profile matches smoothly with the radiative gradient
         beneath.
   \item In the overshoot region $\nabla$ is in between $\nabla_{\rm rad}$
         and $\nabla_{\rm ad}$.
   \item The lower part of the convection zone is weakly subadiabatic.
\end{enumerate}
In this investigation we use a parameterization
which produces overshoot profiles 
possessing the above-mentioned properties.
After finding the overshoot profile that is most consistent with helioseismic data we shall return to overshoot models and discuss potential implications.
}

The idea is to provide an analytical match to the behaviour
obtained in the numerical simulations by Rempel (2004),
constrained such that the temperature gradient and its first derivative
are everywhere continuous.
This is done by representing the actual temperature gradient
as a function of $r$ thus:
\begin{equation}
   \nabla = \nablaad -
   \CFovs {2 \over \beta + \exp(2 \zeta)} \;
   \qquad \hbox{for} \qquad r \ge \rf \; ,
   \label{eq:formul}
\end{equation}
where $\zeta = (r - \rt)/d$.
For $r < \rf$, the temperature gradient is radiative, $\nabla = \nablarad$.
The constants $\CFovs$ and $d$ are determined such that
\begin{equation}
   \left. \begin{array}{l}
   \nabla = \nablarad \\
   \nabla' = \nablarad'
   \label{eq:constraints}
   \end{array}
   \right\} \qquad \hbox{at} \qquad r = \rf \;,
\end{equation}
where the dash indicates differentiation with respect to $r$.
Thus the formulation is characterized by $\rt$, which determines
the overall location of the transition from the adiabatic to radiative temperature
gradient,
and $\rf$, which is the radius at the bottom of the overshoot region,
$\rt {-} \rf$ essentially controlling its width.
We assume the full overshoot region, down to $\rf$, to be
chemically fully mixed.
The parameter $\beta$ provides additional flexibility to the location
of the overshoot region, relative to the base, $r = \rcz$, of
the convectively unstable region, defined by $\nablarad = \nablaad$.

Given $\rt$, $\rf$ and $\beta$, $\CFovs$ and $d$ are determined from
Eqs~(\ref{eq:constraints}), which yield
\begin{equation}
    \CFovs = 
   {1 \over 2} \; \big[\nablaad(\rf) - \nablarad(\rf) \big] \; 
   \big[ \beta + \exp(2 \zetaf) \big] \; ,
   \label{eq:revvalcon}
\end{equation}
where $\zetaf \equiv (\rf - \rt) / d$, and,
neglecting $\nablaad'$,
\begin{equation}
\nablarad' (\rf) 
= - {1 \over d} \;
{2 \exp(2 \zetaf) \over \beta + \exp(2 \zetaf)} \;
\big[\nablarad(\rf) - \nablaad(\rf) \big] \; ,
\label{eq:revdercon}
\end{equation}
on using Eq.~(\ref{eq:revvalcon}).
Equation (\ref{eq:revdercon}) can be solved for $d$
and Eq.~(\ref{eq:revvalcon}) then yields $\CFovs$.

\begin{figure}
   \centering
   \includegraphics[width=\hsize]{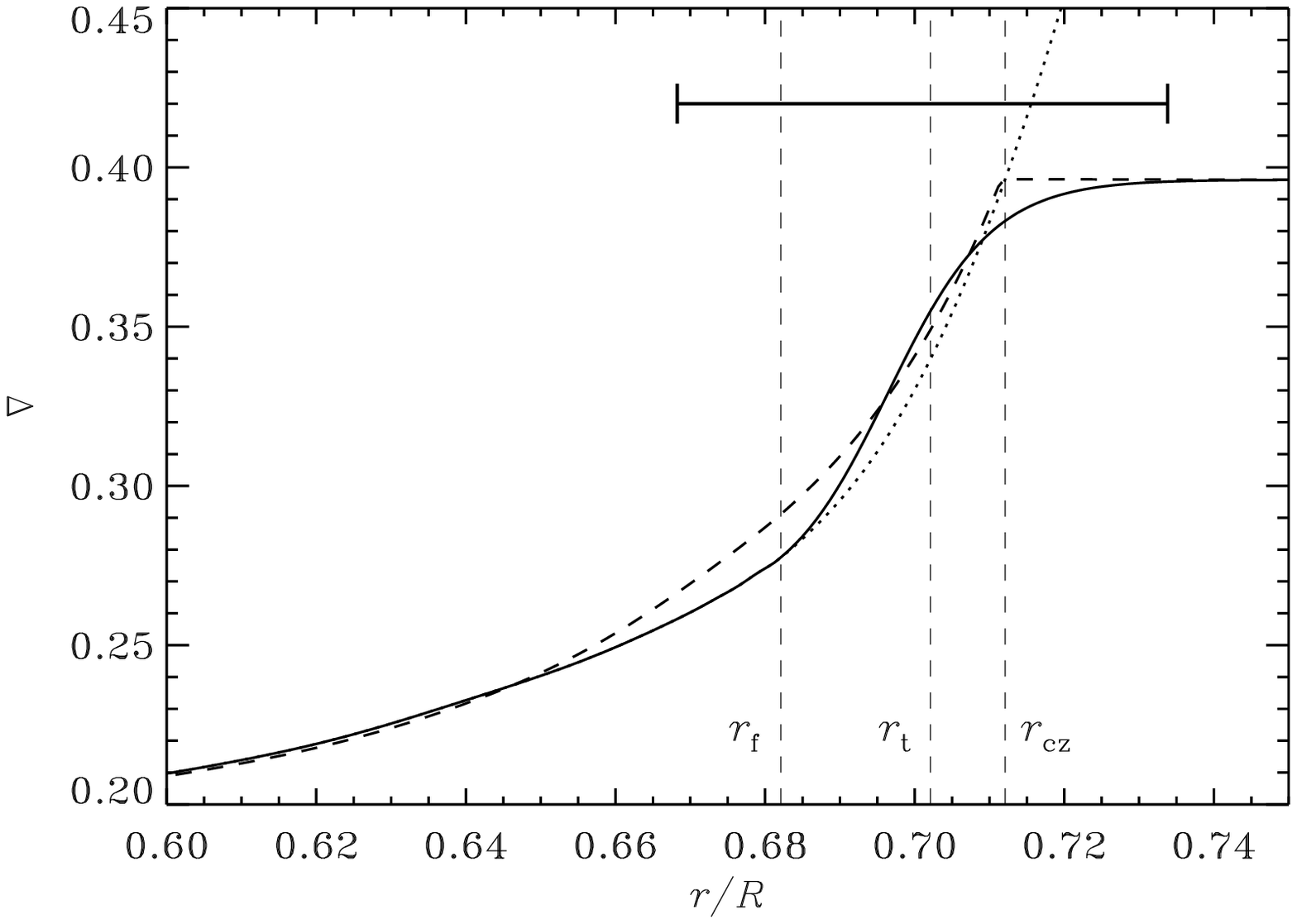}
   \caption{The temperature gradients $\nabla$ (solid curve) and 
$\nablarad$ (dotted curve) in
Model~{\OCtwo} with overshoot characterized by
$\rcz - \rt = 0.01$, $\rt - \rf = 0.02$ and $\beta = 0.5$.
The thin vertical dashed lines show the location of (from left to
right) $\rf$, $\rt$ and $\rcz$.
For comparison, the dashed curve shows $\nabla$ in Model~{\MS} 
of \citet{Christ1996}, with no overshoot.
The horizontal bar shows the range, around $\rt$, over which the argument
$\psi$ (cf.\ Eq. \ref{eq:eigenf}),
at a frequency $\omega/2 \pi = 2500 \muHz$,
changes by $\pi$ and hence provides a measure 
of the vertical wavelength of the acoustic waves.
}
   \label{fig:nabla}
\end{figure}

An example of the resulting $\nabla$ is illustrated in Fig.~\ref{fig:nabla},
compared with the reference Model~{\MS} with no overshoot. 
This clearly illustrates the region of subadiabatic stratification in the 
lower part of the convection zone, and the smooth match to the radiative
gradient at the bottom of the overshoot region, in the model with overshoot.
For comparison, the horizontal bar shows the wavelength of
{\rbf the squared vertical displacement eigenfunction}
$\xi_r^2$ (cf.\ Eq.~\ref{eq:eigenf}) at
a reference frequency $\omega / 2 \pi = 2500 \muHz$;
it is evident that in both models the dominant transition takes place over a
distance substantially smaller than the wavelength.

\section[]{Method of Seismic Analysis}

Following \cite{Monteiro:etal:1994} and \cite{ChristensenDalsgaard:etal:1995},
we consider the frequencies of the modes to be a sum of two components:
a smooth component $\omega_0$,
and an oscillatory component $\delta\omega_{\rm p}$ coming from the 
acoustic glitch caused by the base of the convection zone and any overshoot.

The oscillatory signal arises because of the variation of the phase of the 
mode eigenfunctions at the location of the sharp feature, as a function of 
mode frequency and to a lesser extent of mode degree.
Asymptotically, the radial displacement eigenfunction 
$\xi_r$ is given by
\begin{equation}
\label{eq:eigenf}
\xi_r \ \propto\ (\rho c)^{-1/2} r^{-1} \cos[\psi(r)] \;,
\end{equation}
where 
\begin{equation}
\label{eq:eigphase}
\psi(r) =
\int_r^R \omega \left( 1 - {L^2 c^2 \over \omega^2 r'^2} \right)^{1/2}
{\diff r' \over c} +\phi 
\end{equation}
and $\phi$ is a phase function (it is principally a function of frequency)
which depends on conditions near the 
surface of the Sun: we discuss $\phi$ further below, and in detail in the 
Appendix.

As shown by \citet{ChristensenDalsgaard:etal:1995}, the acoustic
glitch formed by the base of the convection zone and any region of 
overshooting gives rise to an oscillatory signal, with argument
\begin{equation}
 \Lambda(\omega,l) \equiv
    2\omega\btaud - \bgamd\; {l(l{+}1) \over \omega} + 2\phi_0 \;,
    \label{eq:signal-arg}
    \end{equation}
which can be expressed as
 \begin{equation}\begin{array}{l}
 \dis \delta\omega_{\rm p} =
 {1{-}2\Delta \over (1{-}\Delta)^{5/2}} \; a_1
    \left( {\tilde\omega \over \omega} \right)^2 
    \sin [\Lambda(\omega,l)]\\[10pt]
 \dis \qquad
 + {1{-}2\Delta \over (1{-}\Delta)^2} \; a_2
    \left( {\tilde\omega \over \omega} \right) \;
    \cos [\Lambda(\omega,l)] \;.
\label{eq:signal}
\end{array}\end{equation}
The second term is the form expected from a discontinuity in the
first derivative of sound speed, and the first
term from a discontinuity in the second derivative.
Here 
\begin{equation}
\Delta = {l(l{+}1) \over \tilde l (\tilde l {+} 1)}\;
        \left({\tilde\omega \over \omega}\right)^2\; \deld \;
\end{equation}
($\deld$ is defined below) and 
$\tilde l$ and $\tilde \omega$ are reference values of $l$ and $\omega$.
For the present investigation we chose the reference values
$\tilde l {=}20$ and $\tilde\omega{/}2\pi{=}2500\,\mu$Hz.

To obtain Eq.~(\ref{eq:signal}) an expansion in $\Delta$ has been used to 
derive the expression for the amplitude: moreover, it has been
assumed that the acoustic glitch is adequately represented by a single 
location $\rd$ to perform the expansion in frequency.

The phase function $\phi$, due to the reflection of the modes at the inner and outer turning points, is assumed to be represented asymptotically \citep{Monteiro:etal:1994} by
\begin{equation}
\phi(\omega,l) \simeq \phi_0 + a_\phi \omega + a_\gamma {l(l{+}1) \over 2 \omega}\;,
\end{equation}
where $(a_\phi,a_\gamma)$ are unknown expansion coefficients.
The remaining parameters are related to the internal structure of the star
through the following relations:
\begin{eqnarray}
&& \btaud = \int_{\rd}^R {\diff r \over c} + a_\phi \;,
\qquad
\bgamd = \int_{\rd}^R {c \over r^2} \; \diff r \; + a_\gamma \; \; ,
\nonumber \\
&& \deld = {\tilde l ( \tilde l + 1) \over \tilde \omega^2}
\left({c^2 \over r^2}\right)_{r_{\rm d}} ,
\end{eqnarray}
while $a_1$ and $a_2$ depend on the structure at the transition
(see \citealt{ChristensenDalsgaard:etal:1995}, for the details).
Note that the measurable quantity $\btaud$ is not identical to $\taud$,
the acoustic depth of the acoustic glitch, because it includes a 
contribution $a_\phi$ from the dependence of $\phi$ on frequency.
Similarly, $\bgamd$ contains a contribution from the $l$-dependent part
of $\phi$.

In the present work we are mainly interested in exploring the nature of the 
profile of the transition near the base of the convection zone.
Thus we wish to measure the different contributions to the signal coming 
from the discontinuities in the first and second derivatives of the sound speed.

In previous works we have found that if $\phi_0$ is also fitted as a free parameter, there is a strong correlation between the resulting values of
$\btaud$ and $\phi_0$.
This probably results from the fairly narrow range of frequencies
included and the weak dependence on mode degree for most of the points, which means that a change in the average of $\omega \btaud$ can
be compensated by a change in $\phi_0$ and $\deld$.
To overcome this difficulty the two terms in expression (\ref{eq:signal}) were 
{\rbf previously} combined in a single cosine function with a single amplitude that depends on frequency and mode degree.
However, this option weakens our capacity to study the actual behaviour of $a_1$ and $a_2$ for different profiles of the transition.
Thus, in a departure from what we have done previously in 
\cite{Monteiro:etal:1994} and \citet{ChristensenDalsgaard:etal:1995}, here 
we use directly the expression (\ref{eq:signal}) in the fit to the data to 
determine the four parameters
$(a_1,a_2,\btaud,\bgamd)$.
The two components of the signal are fitted in the frequencies of modes with 
degrees $5\le l\le 20$ and cyclic frequencies
$1800 \le \omega/2\pi \le 4000\,\mu$Hz using an iterative least-squares procedure.
The numerical approach is the same as described in 
Appendix~C of \cite{Monteiro:etal:1994}.

   \begin{figure}
   \centering
   \includegraphics[width=8cm]{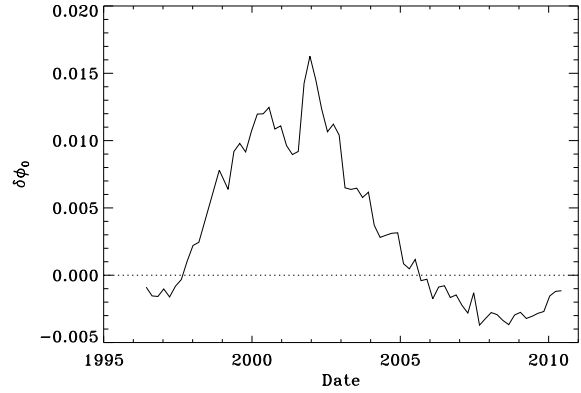}
   \caption{Difference in zero-frequency phase between the observations 
and Model~{\MSpr}, obtained from differential asymptotic analysis
(see Appendix A), plotted against the mean date of the observations.
}
   \label{fig:obsdelphase}
   \end{figure}

As noted above, if $\phi_0$ is fitted as a free parameter it may be highly
correlated with $\btaud$. However,
as discussed in Appendix A, $\phi_0$ can be determined independently
from analysis of the oscillation frequencies
and hence can be regarded as known in the fit to Eq.~(\ref{eq:signal}).
We therefore adopt this procedure.
There is still the difficulty of having $\deld$ as free parameter of the 
fitting, due to the connection with $\bgamd$.
Considering that $\deld$ is a slowly varying function of $\tau$ in all models, 
we have chosen to adopt a reference value from the standard solar model 
corresponding to $\deld=0.35$.

In order to compare the amplitude of the signal for different models and the solar data it is convenient to use a reference value of the amplitude at fixed frequency $\tilde{\omega}$ and at $l=0$ defined by
\begin{equation}
\A25 = \left( a_1^2 + a_2^2 \right)^{1{/}2} \;.
\label{eq:compamp}
\end{equation}

The principal quantities characterizing the signal due to the transition
near the base of the 
convection zone are the period, measured by $\btaud$, and the amplitude (given by $a_1$ and $a_2$).
The secondary quantities $\bgamd$ and $\deld$ are necessary to account for the dependence on mode degree of the amplitude and the period of the signal (see Fig.~3 of \citealt{Monteiro:etal:1994}) as data up to $l=20$ is used.
The advantage of including the higher-degree data is that in general their
frequencies are more precisely determined, and the much higher number of 
mode frequencies being used renders the fitting much more reliable.

Our least-squares fitting procedure weights all data equally,
{\ie}, it does not take the quoted observational uncertainties into account.
The reason for this choice is that we find it gives better determinations of the two amplitude
parameters $a_1$ and $a_2$: otherwise the largely systematic variation in 
mode frequency uncertainties with frequency causes these parameters to 
be less well constrained in the fitting.
In practice it would correspond to putting all the weight on the low-frequency 
range of the spectrum.
This would mainly lead to the determination of one of the $a_i$ coefficients, 
rendering the other unnecessary in the fit.
The solution would then be mostly determined by the initial guess.
Such a frequency-dependent bias of the fitting can be removed if the fitting 
uses equal weights.

\section[]{Analysis of Solar Data}

The solar data used in this work have been obtained from MDI/SOHO 
observations over a total period of over 14 years.
To determine the signal we use sets of frequencies\footnote{The frequencies 
are obtainable online courtesy of J. Schou at \\
{\tt http://quake.stanford.edu/$\sim$schou/anavw72z/}.} 
obtained for 72 days of continuous observations \citep{Schou:1999}.
Sets of frequencies for 1-year periods are also considered.
These have been obtained by combining 5 consecutive sets of 72-d frequencies to produce an average for each 1-year set.
Finally we considered frequencies from a coherently analysed 
set of 1-yr data from the beginning of the MDI observations.
The results from this set were very similar to those of the 
corresponding 1-yr average of the 72-d sets and hence will not
be discussed separately in the following.

We use modes having degree $5\le l \le 20$ and cyclic frequency between 
1900 and $4000\,\mu$Hz. Because the least-squares fitting ignores the 
observational uncertainty on the frequencies, only
modes with a quoted observational uncertainty below $0.1\,\mu$Hz are used.
The phase constant $\phi_0$ was estimated as discussed in Appendix A, from 
a differential asymptotic analysis of each frequency set relative to 
the frequencies of Model~{\MSpr} that was adjusted to match 
approximately the solar frequencies. 
The resulting phase offset $\delta \phi_0$, as a function of the date of the observations, is illustrated in 
Fig.~\ref{fig:obsdelphase}.
In the analysis this offset is added to the phase $\phi_0 = 0.5485$ obtained for
Model~{\MSpr} to obtain the appropriate phase constant for the particular
observational dataset.
The variation of $\delta \phi_0$ with phase in the solar cycle is discussed in
Section~\ref{sec:solarcycle}.

   \begin{figure}
   \centering
   \includegraphics[width=\hsize]{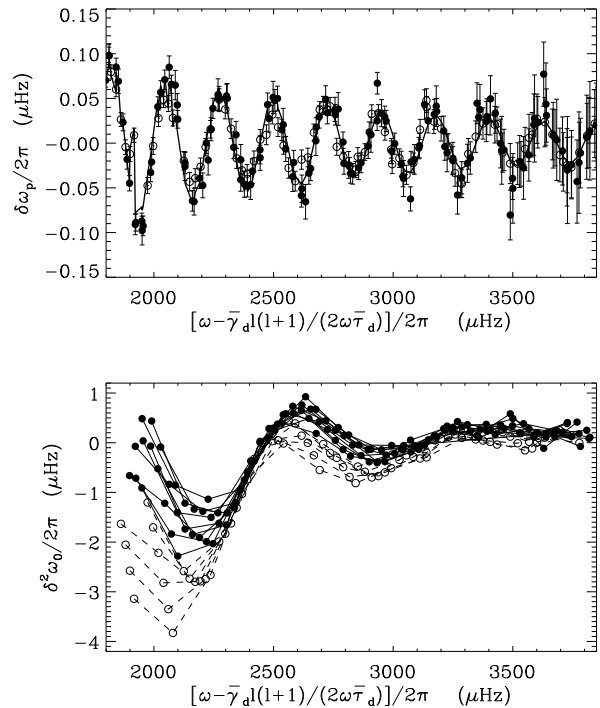}
   \caption{{\em Top panel}: {\rbf Signal for a set of frequencies from MDI (1-year data), shown as filled circles for $5\le \ell \le 14$ and open circles for $15\le \ell \le20$, with $1\,\sigma$ error bars.
Also shown are the fitted values at each point from the expression
in Eq.~(\ref{eq:signal}), 
using the parameters obtained by fitting the frequencies;
to guide the eye, all points have been ordered by the reduced frequency 
used as abscissa and connected by a line.
(The somewhat irregular behaviour arises from the dependence of the 
amplitude on $\omega$ and $l$.)}
      {\em Bottom panel}: Second differences calculated for 
the smooth component $\omega_0$ of the frequencies, obtained by subtracting from the frequencies the values represented by the line in the top panel. The lines connect points of the same mode degree (symbols as above).
}
   \label{fig:signal_sun}
   \end{figure}

   \begin{figure}
   \centering
   \includegraphics[width=\hsize]{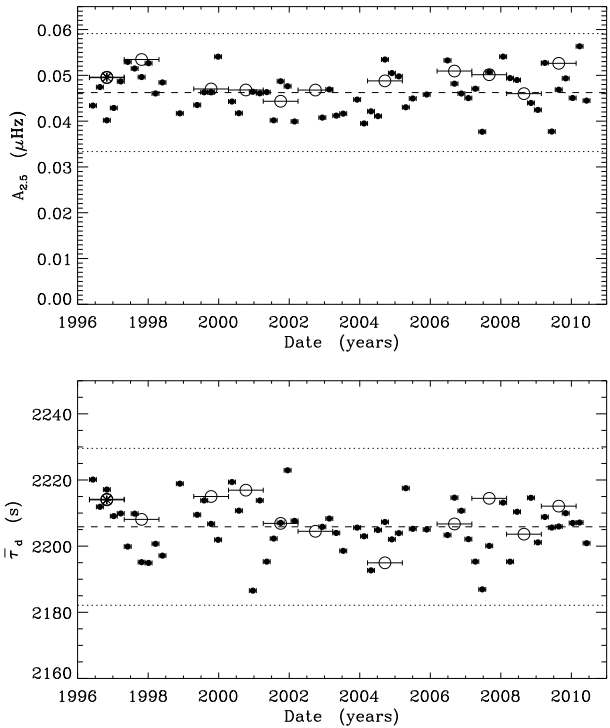}
   \caption{Values of the amplitude $\A25$ (top panel) and acoustic depth $\btaud$ (bottom panel) of the signal for sets of 72 days of MDI frequencies (full circles) and for sets of 1-year MDI frequencies (open circles), plotted against the mean date of the observations.
The horizontal error bar represents the period of the time-series used for the calculation of the frequencies.
      The horizontal dashed line corresponds to the mean,
and the horizontal dotted lines indicate $\pm 3\,\sigma$ standard deviations obtained by Monte Carlo simulations based on 72-d uncertainties of the solar frequencies.}
   \label{fig:date}
   \end{figure}

To illustrate the quality of the fitting,
Fig.~\ref{fig:signal_sun} shows the signal $\delta\omega_{\rm p}$
for the case of frequencies obtained from a 1-year time-series of observations from MDI. 
In order to show that our procedure has cleanly separated the smooth and 
oscillatory components of the frequencies, the lower panel shows
the second differences of the ``smooth'' component of the frequencies, obtained after removing from the solar frequencies the fitted value of the signal at each frequency and mode degree.
This illustrates that the signal coming from the base of the convection zone is no longer visible, while the signal due to the helium ionization zone (e.g. \citealt{Monteiro:Thompson:2005}) can be clearly seen in the second differences.

\begin{table}
\begin{center}
 \caption{Seismic parameters (averages) resulting from the fit of the signal in Eq.~(\ref{eq:signal}) to the solar frequencies.
 Data sets for 72 days have been considered,
as well as a coherently analyzed 360-d data set together with 1-year averages of 5 consecutive sets of 72-d frequencies.
 The values $\sigma_{\rm obs}$ are obtained as the $1\,\sigma$ distribution of the values found for each group of sets of solar frequencies (72-d and 360-d) while $\sigma_{\rm MC}$ are the $1\sigma$ of the Monte Carlo simulation (500 realizations) for a single set of frequencies using the quoted observational uncertainties in the frequencies for one of the 72-d sets and for one of the 1-year sets.
\label{tab:sun}
}
\begin{tabular}{rcccrr}
\hline\hline\\[-6pt]
Type \,\,\, & $\A25$ & $a_1$ & $a_2$ & $\bar\tau_d$ & $\bgamd$ \,\, \\
  & ($\mu$Hz) & ($\mu$Hz) & ($\mu$Hz) & (s) & ($\mu$Hz) \\[2pt]
\hline\\[-6pt]
$\langle$72-d$\rangle$ & 0.0462 & 0.0442 & 0.0101 & 2206 & 16.7 \\
$\sigma_{\rm obs,72}$ & 0.0045 & 0.0045 & 0.0088 & 8 & 2.3 \\[5pt]
$\langle$360-d$\rangle$ & 0.0488 & 0.0471 & 0.0095 & 2209 & 17.5 \\
$\sigma_{\rm obs,360}$ & 0.0027 & 0.0042 & 0.0079 & 7 & 1.4 \\
\hline\\[-8pt]
$\sigma_{\rm MC,72}$ & 0.0043 & 0.0045 & 0.0087 & 8 & 2.1\\
$\sigma_{\rm MC,360}$ & 0.0019 & 0.0018 & 0.0010 & 2 & 0.8\\[2pt]
\hline\hline
\end{tabular}
\end{center}
\end{table}

The impact of the observational uncertainties on $\btaud$, $\A25$
and the other measured parameters has been estimated in two ways.
Firstly we compute the standard deviations of the measured values for
a given parameter amongst all the 72-d observational datasets.
Secondly, we make 500 Monte Carlo simulations of frequency datasets,
using the estimated errors for a set of 72-d observational frequencies,
fitting those artificial datasets
in the same way as we fit the real data, and calculate standard deviations 
for the resulting estimated parameters.
The estimate based on all the observed frequency sets 
will be affected by any temporal variations in the solar 
values, since the data are obtained at different epochs.
The second estimate is purely
an indicator of how the uncertainties in the frequencies impact the 
parameters measured from the signal.
Both sets of uncertainties are shown in Table~\ref{tab:sun} together with 
the average values for the Sun.
It is evident that the standard deviations in the parameters inferred
from the observations are largely
consistent with the result of the Monte Carlo simulation.
Thus the scatter in the inferred values for $\A25$ 
for the observations is essentially consistent with the assumed
error in the observed frequencies.

Figure~\ref{fig:date} shows the inferred amplitude $\A25$ and period $\btaud$
for the solar data, as functions of epoch.
The dashed lines indicate the average values over the whole period,
and the dotted lines indicate the $3\,\sigma$ error interval obtained from
Monte Carlo simulations based on the errors in the 72-d datasets.
When 1-year data are used (see Table~\ref{tab:sun}),
there is some indication that the variation in the Sun 
in $\btaud$ and $\bar\gamma_{\rm d}$ is slightly above the noise of the data. 
This could point towards a variation in time of the stratification 
of the overshoot region or the near-surface layers; 
there is indeed perhaps some hint of
an orderly systematic variation of the 1-yr results in Fig.~\ref{fig:date}.
(although not at an 11-yr period),
but more accurate data would be necessary to confirm this possibility.

   \begin{figure}
   \centering
\includegraphics[width=\hsize]{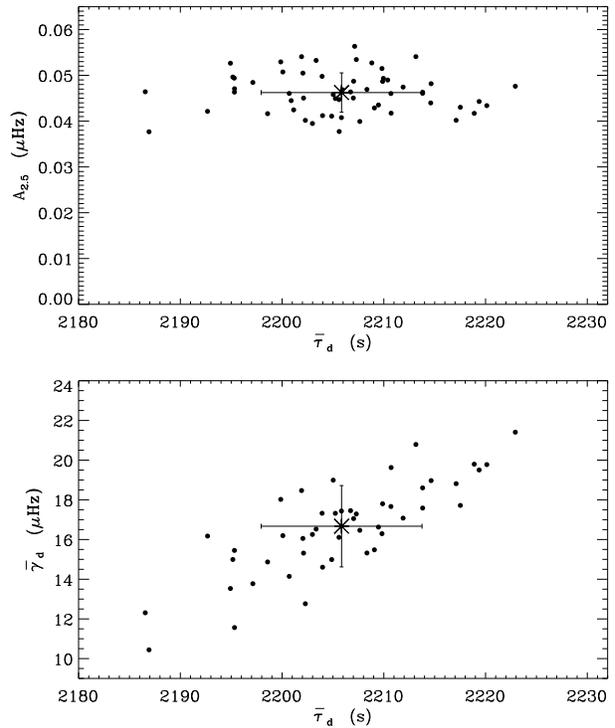}
\caption{
The filled circles are the values of $\A25$ (upper panel) and $\bar \gamma_{\rm d}$ (lower panel), versus $\bar \tau_{\rm d}$, for the solar data (72-d frequency sets).
Also shown, as stars, are the average values of these values together with the error bars corresponding to a $1\,\sigma$ uncertainty estimated from Monte Carlo simulations for the uncertainties of a 72-d set of solar frequencies (see Table~\ref{tab:sun}).}
\label{fig:obsprop}
\end{figure}

As a further indication of the properties of the parameters inferred from
the 72-d datasets, 
Fig.~\ref{fig:obsprop} shows $\A25$ and $\bar \gamma_{\rm d}$ 
against $\btaud$.
While $\A25$ and $\bar \tau_{\rm d}$ show no obvious correlation,
there is a clear correlation between $\btaud$ and $\bgamd$.
In fact it may be seen from Eq.~(\ref{eq:signal}) that both $\btaud$ and
$\bgamd$ affect the phase of the signal, in such a way that an increase
in $\btaud$ can be compensated by an increase in $\bgamd$, as observed in the lower panel of Fig.~\ref{fig:obsprop}.
This correlation is also confirmed by Monte Carlo simulations carried out
on the basis of the model frequencies (see below). 
For other pairs of parameters no significant correlation was found.

\section[]{Analysis of solar models}

The signal for the models has been fitted using the same range of mode degree and frequency as done for the Sun (see above).
In this way we have ensured that the fitting takes place under the same conditions as for the solar data, both for error free data and for the Monte Carlo simulations used to estimate the impact of noise.

\subsection[]{Models emulating overshoot} \label{sec:models}

Apart from the treatment of the overshoot region, the models considered 
here were computed in the same way as Model~S of \citet{Christ1996}.
The computation used the OPAL equation of state \citep{Rogers1996} and
opacity \citep{Iglesi1992},
the \citet{Bahcal1995} nuclear reaction rates, and the \citet{Michau1993}
treatment of diffusion and settling of helium and heavy elements, the
latter being treated as fully ionized oxygen.
All models were calibrated, 
to within a relative accuracy of $10^{-6}$, to a photospheric radius of 
$6.9599 \times 10^{10} \, {\rm cm}$, a surface luminosity of
$3.846 \times 10^{33} \, {\rm erg \, s^{-1}}$ and a ratio
$Z_{\rm s}/X_{\rm s} = 0.0245$ between the surface abundances by mass of
heavy elements and hydrogen, at a model age of 4.6 Gyr.
Further details on the model and oscillation calculations were given
by \citet{Christ2008a, Christ2008b}.

{\rbf To explore the sensitivity of the oscillations to the structure of the
overshoot region we have considered a range of models with various
characteristics, including models exhibiting the `classical' sharp overshoot
profile, a sequence of models converging towards the reference model,
as well as models exhibiting various degrees of smoothness in the transition
of the temperature gradient.}
Properties of the models are provided in Table~\ref{tab:models}.
In discussing the models with overshoot or otherwise modified, we use
Model~{\MS} as reference.

\begin{table*}
\caption{Model properties.
Here {\MS} refers to Model~S of \citet{Christ1996}, {\MSpr} to a similar
model, but with modified $\gamma_1$ in the near-surface layers
(cf.\ Eq.~\ref{eq:gammod}), and models {\OAone} -- {\OCthree} were obtained
utilizing the overshoot formulation described in Section~\ref{sec:param},
and characterized by $\rt$, $\rf$ and $\beta$, $r_{\rm cz}$ and
$\tau_{\rm cz}$ being the radius and acoustic depth at
the base of the convection zone, defined by $\nablarad = \nablaad$.
Model~{\OD} in addition had a modest amount of turbulent mixing beneath the
overshoot region while in Model~{\OP} the opacity near the boundary of the
unstable region was reduced slightly (see text).
To characterize further the properties of the models,
$\tau_{\rm max}$ shows the acoustic depth at the maximum in
the overshoot region of $\diff \nabla /\diff r$,
and $\ell_{\rm ov}/H_p$ is the total extent $\rcz - \rf$ of the overshoot
region, in units of the pressure scale height $H_p$ at the base of the
convection zone.
Finally, $\phi_0$ is the phase at zero frequency, determined from
a fit to the eigenfunctions (cf.\ Appendix A).
\label{tab:models}}
\begin{tabular}{lllccccc}
\hline\hline\noalign{\smallskip}
Model & $\displaystyle {\rcz {-} \rt \over R}$ 
  & $\displaystyle{\rt {-} \rf \over R}$ & $\beta$ 
  & $\begin{array}{c}\tau_{\rm cz} \\ ({\rm s})\end{array}$ 
  & $\begin{array}{c}\tau_{\rm max} \\ ({\rm s})\end{array}$
  &  $\displaystyle{\ell_{\rm ov}\over H_p}$ &  $\phi_0$ 
\\[5pt]
\hline\noalign{\smallskip}
\MS     & -     & -     & -   & 2171 & 2176 &   -  & 0.8904 \\ 
\MSpr   & -     & -     & -   & 2176 & 2181 &   -  & 0.5485 \\ 
\OAone  & 0.01  & 0.001 & 1.0 & 2170 & 2204 & 0.14 & 0.8916 \\ 
\OAtwo  & 0.005 & 0.001 & 1.0 & 2171 & 2189 & 0.07 & 0.8914 \\ 
\OBone  & 0.01  & 0.005 & 1.0 & 2170 & 2203 & 0.19 & 0.8898 \\ 
\OBtwo  & 0.005 & 0.0025& 1.0 & 2170 & 2188 & 0.09 & 0.8902 \\ 
\OBthree& 0.0025& 0.0010& 1.0 & 2171 & 2181 & 0.04 & 0.8903 \\ 
\OCone  & 0.01  & 0.015 & 0.5 & 2168 & 2212 & 0.31 & 0.8938 \\ 
\OCtwo  & 0.01  & 0.02  & 0.5 & 2170 & 2217 & 0.37 & 0.8944 \\ 
\OCthree& 0.01  & 0.025 & 0.5 & 2170 & 2219 & 0.44 & 0.8951 \\ 
\OD     & 0.01  & 0.02  & 0.5 & 2167 & 2217 & 0.37 & 0.8957 \\ 
\OP     & 0.02  & 0.02  & 1.0 & 2164 & 2227 & 0.50 & 0.8954 \\ 
\noalign{\smallskip}
\hline\hline
\end{tabular}
\end{table*}

   \begin{figure}
   \centering
   \includegraphics[width=\hsize]{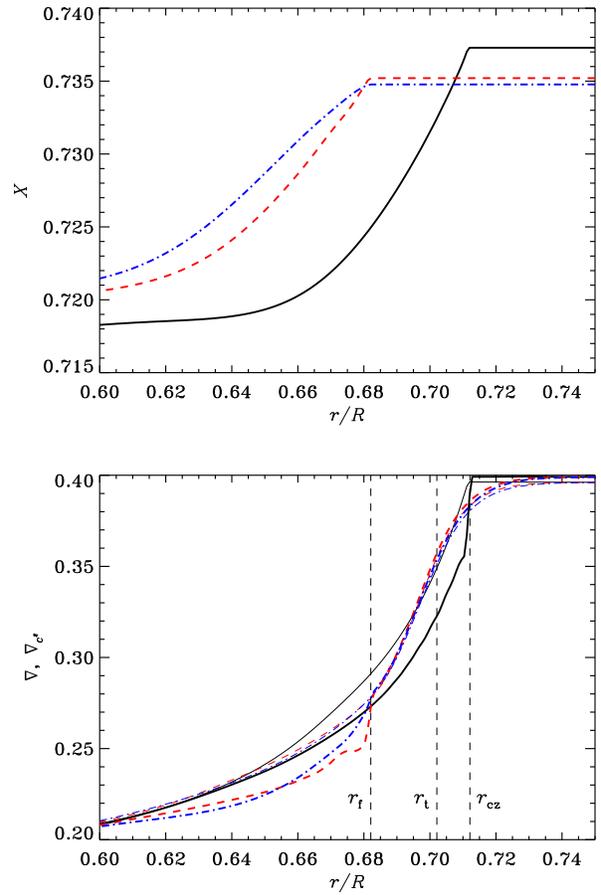}
   \caption{
Properties of reference Model~{\MS} (black solid lines), overshoot Model~{\OCtwo}
(red dashed lines) and Model~{\OD} (blue dot-dashed line)
which in addition has turbulent diffusion below the overshoot
region.
{\it Top panel}: Hydrogen abundance $X$ against fractional radius $r/R$.
{\it Bottom panel}: The thin curves show the temperature gradient
$\nabla = {\dd \ln T / \dd \ln p}$, and the heavier curves the corresponding
gradient in the squared sound speed (cf.\ Eq. \ref{eq:nablacsq}).
{\rbf As in Fig.~\ref{fig:nabla} the vertical dashed lines mark the location
of $\rf$, $\rt$ and $\rcz$ in Model~{\OCtwo}.}
}
   \label{fig:nabcsq}
   \end{figure}


Although the properties of the overshoot region are described in terms of
the temperature stratification, the effect on the oscillation frequencies
is predominantly controlled by the sound speed which in addition is 
affected by the composition. Of particular importance is
the relatively steep gradient
in the hydrogen abundance ($X$) established just below the fully mixed region
which, as mentioned above, includes the overshoot region down to $r = \rf$.
This affects the sound-speed gradient
\begin{equation}
\nablacsq \equiv {\dd \ln c^2 \over \dd \ln p} 
\simeq \nabla - {\dd \ln \mu \over \dd \ln p}\; ,
\label{eq:nablacsq}
\end{equation}
where the last approximation used Eq.~(\ref{eq:apprcsq}).
This behaviour is illustrated in Fig.~\ref{fig:nabcsq}.
It is evident that the gradient in $X$ accentuates the gradient
in $c^2$ in Model~{\MS} without overshoot \citep[e.g.,][]{Basu1994},
and similarly produces a step in $\nablacsq$ at the edge of the
overshoot region in Model~{\OCtwo} illustrated in the figure.

One might argue that further motion beyond the overshoot region would cause
additional diffusive mixing, thus likely smoothing out the step in the
sound-speed gradient.
To investigate the effect of this on the signal we have computed Model~{\OD},
corresponding to Model~{\OCtwo} but
with additional diffusive mixing in a small region beneath $\rf$;
specifically, the maximum diffusion coefficient, at the edge of the
overshoot region, was $50\, {\rm cm^2/s}$,
decreasing rapidly with depth.
This causes a modest smoothing of the hydrogen profile which evidently 
is sufficient to eliminate the step in $\nablacsq$.

   \begin{figure}
   \centering
   \includegraphics[width=\hsize]{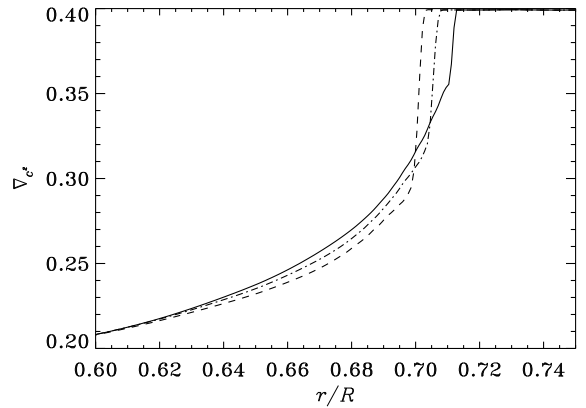}
   \caption{
Gradients $\nablacsq$ of the squared sound speed in the reference Model~{\MS}
(solid line) and Models~{\OAone} (dashed line) and {\OAtwo} (dot-dashed line)
with a sharp transition at the bottom of the overshoot region.
}
   \label{fig:nabsharp}
   \end{figure}

   \begin{figure}
   \centering
   \includegraphics[width=\hsize]{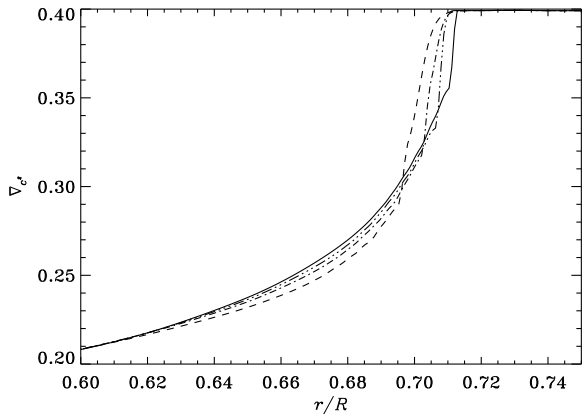}
   \caption{
Gradients $\nablacsq$ of the squared sound speed in the reference Model~{\MS}
(solid line) and Models~{\OBone} (dashed line), {\OBtwo} (dot-dashed line)
and  {\OBthree} (triple-dot-dashed line) converging to Model~{\MS}
(see also Table~\ref{tab:models}).
}
   \label{fig:nabseq}
   \end{figure}

   \begin{figure}
   \centering
   \includegraphics[width=\hsize]{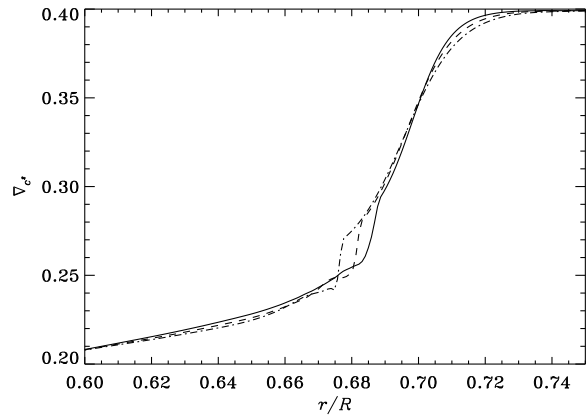}
   \caption{
Gradients $\nablacsq$ of the squared sound speed in 
Models~{\OCone} (solid line), {\OCtwo} (dashed line) and
{\OCthree} (dot-dashed line), illustrating increasingly smooth overshoot.
}
   \label{fig:nabsmooth}
   \end{figure}

   \begin{figure}
   \centering
   \includegraphics[width=\hsize]{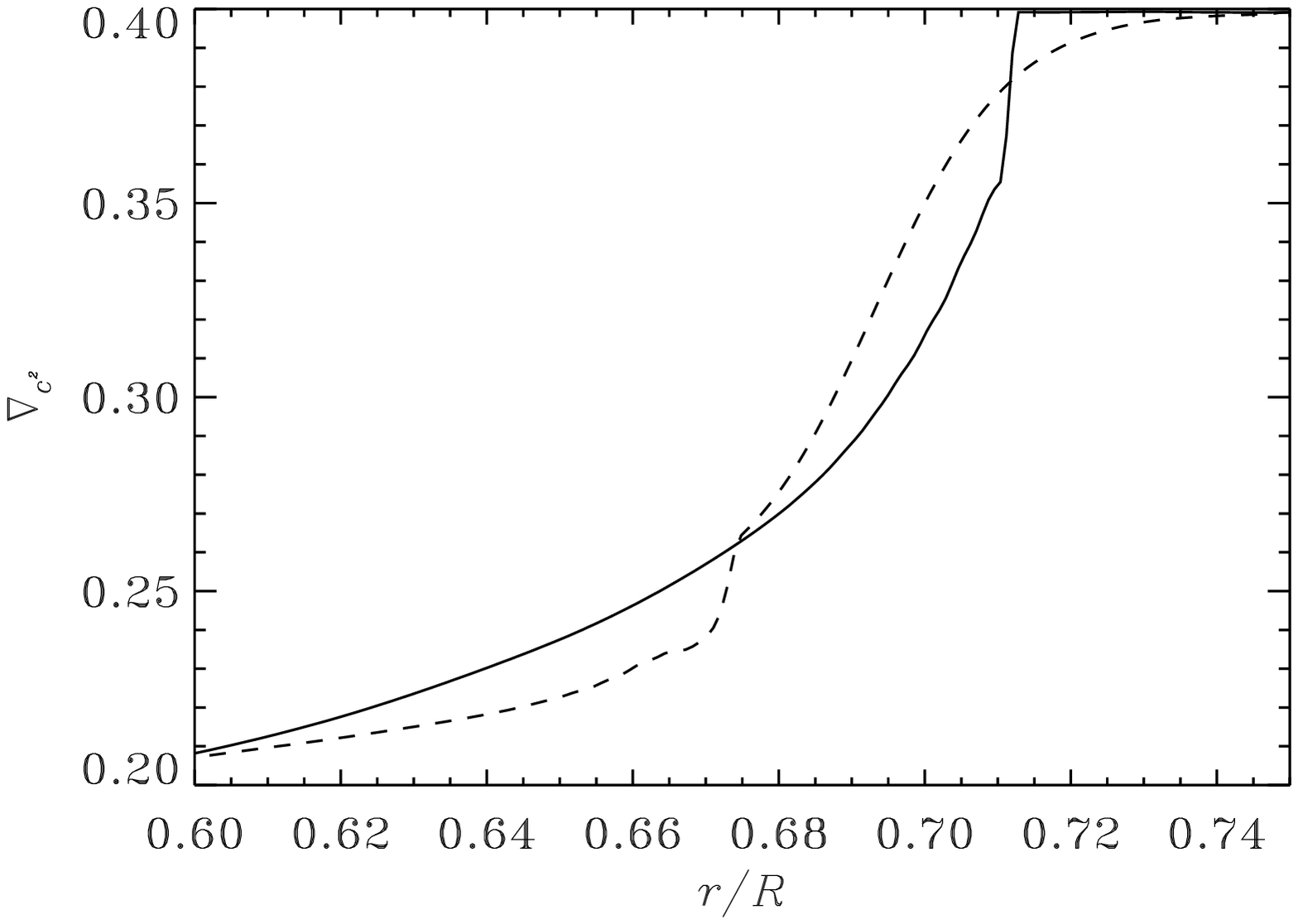}
   \caption{
Gradients $\nablacsq$ of the squared sound speed in the reference Model~{\MS}
(solid line) and Model~{\OP} (dashed line), with a very smooth transition
and a small opacity reduction near the base of the convection zone.
}
   \label{fig:nabvsmooth}
   \end{figure}


As a further background for the discussion of the model signals below,
we first present a range of examples of profiles of the sound-speed gradient 
obtained with the formalism described in Section \ref{sec:param}.
These were chosen to provide an impression of the sensitivity of the signal to
the detailed structure near the base of the convection zone, as well
as to illustrate the structure that may be consistent with the observations.
The input parameters of the models, and some relevant characteristics, 
are listed in Table~\ref{tab:models}.
Here, $\tau_{\rm cz}$ is the acoustic depth of the boundary $\rcz$ of 
the convectively unstable region, while $\tau_{\rm max}$ is the acoustic depth
of the point of steepest gradient in $\nabla$, providing an indication of
the location of the acoustic glitch.
Also, $\phi_0$ is the extrapolated near-surface phase at zero frequency,
determined from a fit to the eigenfunctions (see Section \ref{sec:eigphase}).
Finally, for comparison with the common characterization of overshoot,
$\ell_{\rm ov}/H_p$ shows the overshoot distance $\rcz - \rf$, in units
of the pressure scale height at $\rcz$.
The properties of the models are illustrated in Figs~\ref{fig:nabsharp} -- 
\ref{fig:nabvsmooth} in terms of $\nablacsq$.

Figure~\ref{fig:nabsharp} shows that the present formulation 
emulates the sharp transition predicted by non-local mixing-length theories
\citep[e.g.,][]{Zahn1991},
and discussed in earlier work
\citep{Basu:etal:1994, Basu1994, Monteiro:etal:1994,
Roxburgh:Vorontsov:1994, ChristensenDalsgaard:etal:1995}
which provided stringent constraints on overshoot of this form.
Figure \ref{fig:nabseq} shows a sequence of models converging to Model~{\MS}
(see also the parameters in Table~\ref{tab:models}).
A sequence of models of gradually increasing smoothness is illustrated
in Fig.~\ref{fig:nabsmooth}.
Finally, Fig.~\ref{fig:nabvsmooth} shows a model with a very smooth transition.
To ensure a reasonable sound-speed profile, in view of the helioseismic
inference (see also Section \ref{sec:compmod}), we shifted the boundary $\rcz$
of the unstable region
outwards by about $0.002 R$ relative to Model~{\MS}, through a 
localized reduction of the opacity by 7 per cent,
centred near the bottom of the convective envelope.

The frequencies of all models have been fitted with Eq.~(\ref{eq:signal}) in order to obtain the four parameters
$(a_1,a_2,\btaud,\bgamd)$.
Over 235 frequencies for each model have been used in the fit and no 
noise has been added to the frequencies.
The values of the parameters found are listed 
in the first columns of Table~\ref{tab:sigmod}.

\begin{table*}
\caption{Seismic parameters resulting 
from the fit of the signal in Eq.~(\ref{eq:signal}) to the frequencies 
of models.
The first three columns are the results for noise-free data
while the other six columns provide the results for 500 Monte Carlo simulations of each model when the uncertainties from a solar 72-d set of frequencies 
are considered.
\label{tab:sigmod}}
\begin{tabular}{p{1.5cm}ccp{1.5cm}cccccc}
\hline\hline\noalign{\smallskip}
Model \qquad & $\A25$  & $\bar\tau_d$ & $\hspace{2mm}\bgamd$
   & ${\langle}\A25{\rangle}$ & $\sigma(\A25)$ \qquad
   & ${\langle}\btaud{\rangle}$ & $\sigma(\btaud)$ \qquad
   & ${\langle}\bgamd{\rangle}$ & $\sigma(\bgamd)$ \\
   & ($\mu$Hz) & (s) & ($\mu$Hz) & ($\mu$Hz) & ($\mu$Hz) & (s) & (s) & ($\mu$Hz) & ($\mu$Hz) \\[5pt]
\hline\noalign{\smallskip}
\MS       & 0.0673 & 2164 & 13.1  & 0.0615 & 0.0043  & 2171 & 6  & 16.1 & 1.7 \\
\OAone    & 0.1036 & 2184 & 13.5  & 0.0887 & 0.0046  & 2189 & 5  & 16.1 & 1.5\\
\OAtwo    & 0.0872 & 2173 & 13.4  & 0.0767 & 0.0043  & 2180 & 5 & 16.2 & 1.5 \\
\OBone    & 0.0950 & 2185 & 13.6  & 0.0824 & 0.0048  & 2191 & 5  & 16.4 & 1.6 \\
\OBtwo    & 0.0852 & 2175 & 13.4  & 0.0753 & 0.0045  & 2182 & 5  & 16.4 & 1.5 \\
\OBthree  & 0.0776 & 2169 & 13.2  & 0.0698 & 0.0046  & 2177 & 5  & 16.3 & 1.6 \\
\OCone    & 0.0660 & 2193 & 14.0  & 0.0604 & 0.0050  & 2198 & 6  & 16.7 & 2.0 \\
\OCtwo    & 0.0512 & 2199 & 14.1  & 0.0484 & 0.0047  & 2204 & 8  & 17.4 & 2.7 \\
\OCthree  & 0.0373 & 2201 & 14.0  & 0.0378 & 0.0045  & 2208 & 11  & 18.6 & 3.8 \\
\OD       & 0.0410 & 2192 & 13.5  & 0.0387 & 0.0044  & 2200 & 10 & 17.4 & 3.5 \\
\OP       & 0.0314 & 2208 & 14.1  & 0.0330 & 0.0045  & 2215 & 11  & 19.1 & 4.0 \\
\noalign{\smallskip}
\hline\hline
\end{tabular}
\end{table*}

   \begin{figure}
   \centering
   \includegraphics[width=\hsize]{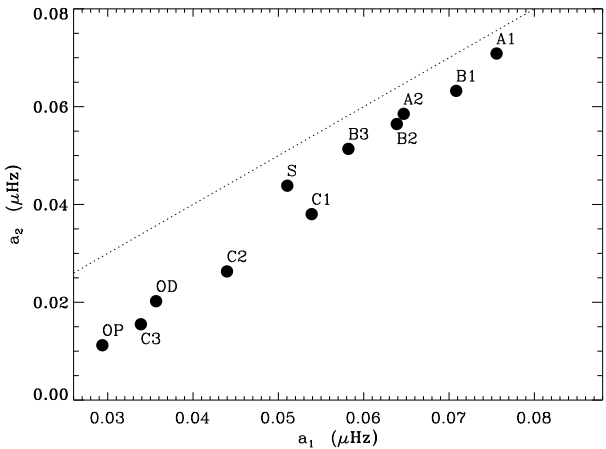}
   \caption{
   The stars show the values of the amplitude parameters $a_1$ and $a_2$
of the signal for all models listed in Table~\ref{tab:sigmod}.
The dotted line corresponds to $a_1=a_2$.
}
   \label{fig:amps-mods}
   \end{figure}
   
   \begin{figure}
   \centering
   \includegraphics[width=\hsize]{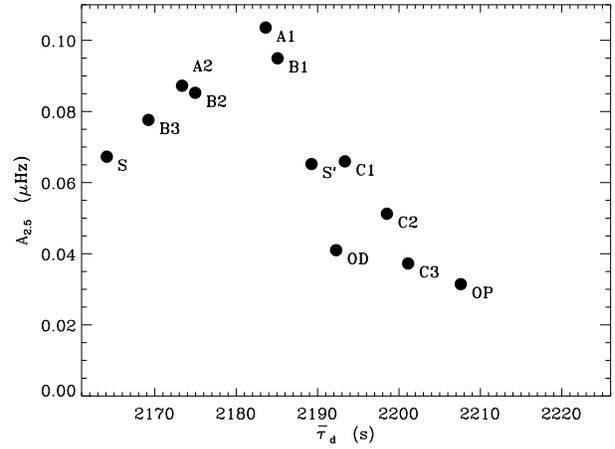}
   \caption{
   The symbols show the values of the amplitude $\A25$
versus acoustic depth $\btaud$ of the signal for all models listed in
Table~\ref{tab:models}, obtained by fitting the noise-free frequencies (values listed in Table~\ref{tab:sigmod}).
}
   \label{fig:signal-mods}
   \end{figure}

The values for $\A25$ (see Eq.~\ref{eq:compamp})
were obtained from the fitted parameters $a_1$ and 
$a_2$ which, though not listed in the table, are illustrated in 
Fig.~\ref{fig:amps-mods}.
The amplitude values depend mainly on the local conditions at the base 
of the convection zone and are insensitive to changes in the model that 
do not affect this layer.
Therefore the distribution of the models in Fig.~\ref{fig:amps-mods} is an accurate representation of the sharpness of the transition at the base of the envelope.
We note in particular that, formally, $a_2$ arises from a discontinuity in the 
first derivative in the sound speed, while $a_1$ arises from a discontinuity
in the second derivative.
Thus one would expect $a_2$ to be bigger relative to $a_1$ for sharper
transitions, as is indeed the case, although both terms contribute for
all the cases considered.
The corresponding inferred values of the amplitude $\A25$ for the models 
are shown in Fig.~\ref{fig:signal-mods}.

These results confirm what was already found by \cite{ChristensenDalsgaard:etal:1995}: in order to have an amplitude below what is found for the standard model {\MS}, a sub-adiabatic region within the proper convection zone (above the Schwarszchild boundary) is required.

\subsection{Impact of noise on the seismic parameters}

When noise is added, the inferred values of the parameters are affected.
The values in the rightmost columns of Table~\ref{tab:sigmod} were obtained from fitting model data after adding solar noise.
For each model, Monte Carlo simulations were run with 500
realizations of errors characteristic of a 72-d solar 
observational dataset.

   \begin{figure}
   \centering
   \includegraphics[width=\hsize]{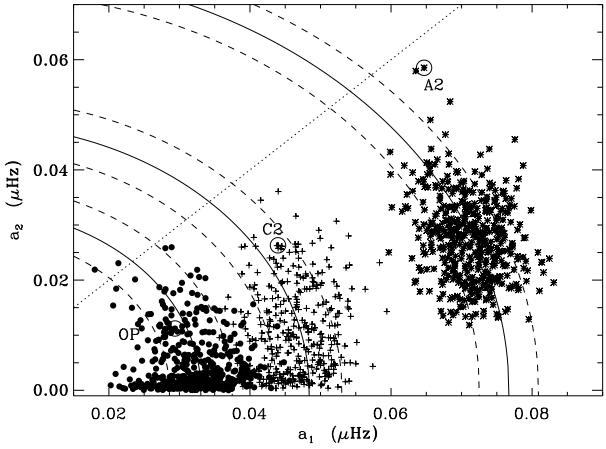}
   \caption{
   Examples for three models ({\OAtwo} - stars; {\OCtwo} - crosses; {\OP} - filled circles) of the values $a_2$ and $a_1$ obtained from Monte Carlo simulations using solar uncertainties from a 72-d set of frequencies.
The solid and dashed curves indicate the corresponding average combined
amplitude $\A25$ (cf.\ Eq.~\ref{eq:compamp}) and the $\pm 1\,\sigma$ error
interval on $\A25$.
The values $(a_1, a_2)$ obtained for the noise-free frequencies are also
indicated as open circles for the three models (see Table~\ref{tab:sigmod}).
}
   \label{fig:aamodmc}
   \end{figure}

Figure~\ref{fig:aamodmc} illustrates the effect on the individual amplitudes $a_1$ and $a_2$ for three models.
It is clear from the figure that the added noise causes
a systematic shift in the inferred values of $a_1$ and $a_2$, with $a_2$ 
tending to be shifted to lower values, essentially corresponding to an apparently smoother transition. 
Similarly, although the value of $\A25$ is more robust, 
it suffers a reduction which, for the sharper transitions,
may exceed $2\,\sigma$: this is shown in Table~\ref{tab:sigmod} and is apparent for Model~{\OAtwo} in Fig.~\ref{fig:aamodmc}.
Even so, the results confirm that, when properly calibrated for the presence of noise, the value of $\A25$ can be used to probe the sharpness of the base of the convection zone.

   \begin{figure}
   \centering
   \includegraphics[width=\hsize]{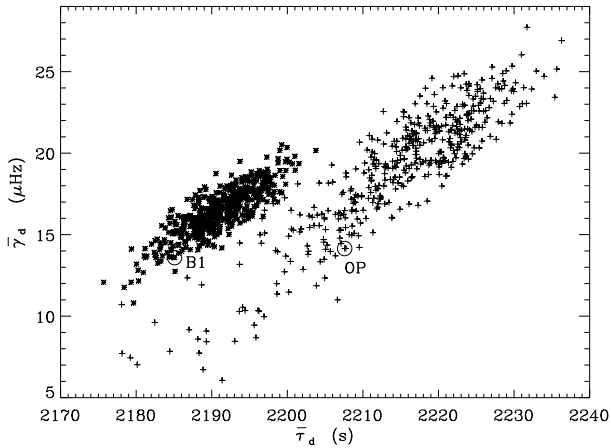}
   \caption{
Values of $\btaud$ and $\bgamd$ for two models, {\OBone} and {\OP}, inferred from the Monte Carlo simulation, based on observed errors.
Note the strong correlation between these two parameters, already shown for the solar data in Fig.~\ref{fig:obsprop}.
The open circles show the results of fitting the noise-free frequencies.
}
   \label{fig:modcor}
   \end{figure}
   
As is the case for solar data, the largest correlation between inferred parameters is that between
$\btaud$ and $\bgamd$: the effect of noise on these two parameters for model
data is shown in Fig.~\ref{fig:modcor}. The standard deviations of the 
inferred values of the parameters, and their mean values, are listed in 
Table~\ref{tab:sigmod}. 
In this table, the values found for the seismic parameters in the presence of observed uncertainties are listed in the last 6 columns, together with the $1\,\sigma$ uncertainties obtained through Monte Carlo simulations of 500 realizations.
These use the uncertainties for a 72-d set of solar frequencies reported before.

As one would expect, the precision of the fitting is degraded when the amplitude is lower.
In particular for $\btaud$ and $\bgamd$, as lower amplitude decreases the ability to separate the frequency and mode degree contributions for the argument of the signal, leading to a wider variation, within the correlation found, of these two parameters (see also Fig.~\ref{fig:modcor}).

We also note that, contrary to our previous work
\citep{Monteiro:etal:1994,ChristensenDalsgaard:etal:1995},
the option to estimate a priori the value of $\phi_0$ and fix the value of $\deld$ has resulted in an increase of the expected precision of the fitting for the three quantities listed in Table~\ref{tab:sun} ($a_1$, $a_2$, $\btaud$),
required for testing the stratification of the overshoot layer.
This improvement is crucial in securing a proper calibration of the key parameters associated with the sharpness and location of the base of the convection zone.
The increase in precision comes at the expense of systematic shifts between
the noise-free fits and the averages of the model fits including noise, as
is evident in Table~\ref{tab:sigmod} as well as Figs~\ref{fig:aamodmc} and 
\ref{fig:modcor}.
To correct for this the following comparison between the observations and the
model results is based on the outcome of the Monte Carlo simulations for 
the latter.



%

\section[]{Results and discussion}

Based on the comparison of the fitting of the signal for the models
(particularly
in the presence of noise) and for the Sun, it is now possible to establish what are the implications for the stratification of the transition layer at the base of the convection zone.

   \begin{figure}
   \centering
   \includegraphics[width=\hsize]{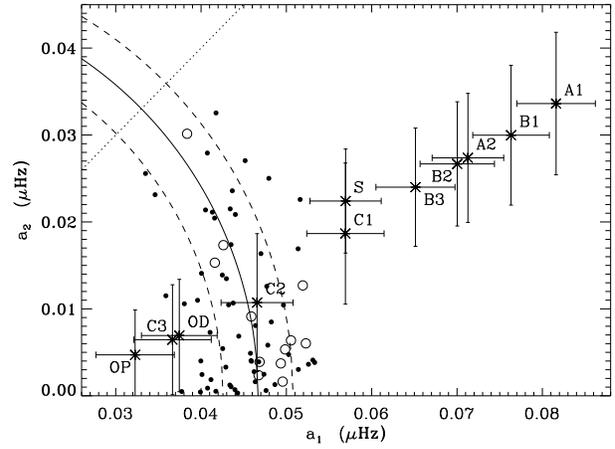}
   \caption{
The stars show the average value from Monte Carlo simulations for the models using solar uncertainties, while the error bars are $1\,\sigma$ errors associated with those uncertainties, as given in Table~\ref{tab:sigmod}.
The dotted line corresponds to $a_1=a_2$.
The small filled circles show the fits to the individual observed 72-d datasets while the open circles are for 1 year datasets.
The dashed lines show the $\pm 1\,\sigma$ range of $\A25$ around the mean value for the solar data (full line).
}
   \label{fig:amps-obs}
   \end{figure}

\subsection{Comparing the models with the Sun} \label{sec:compmod}

\begin{figure}
   \centering
   \includegraphics[width=\hsize]{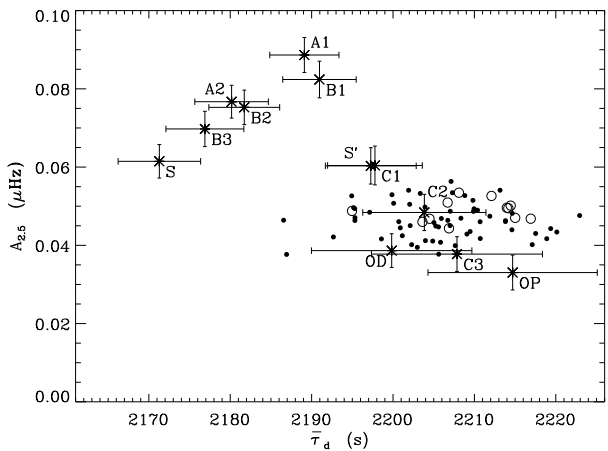}
   \caption{
The small filled circles show the fits to the individual 72-d observed datasets while open circles are for 1 year datasets.
The stars show the average values obtained for Monte Carlo simulations of the models with $1\,\sigma$ error bars, as listed in Table~\ref{tab:sigmod}.
}
   \label{fig:signal-obs}
   \end{figure}

Figure~\ref{fig:amps-obs} compares the amplitudes ($a_1$ and $a_2$) obtained
from solar data with those obtained from model frequencies with solar-like noise added 
(cf.\ the noise-free case in Fig.~\ref{fig:amps-mods}).
The model that seems most consistent with the solar observations, in this
representation, is Model~{\OCtwo}.
This is confirmed by a comparison of the inferred 
values of $\A25$ and $\btaud$ for solar data and noisy model data
(Fig.~\ref{fig:signal-obs}; 
cf.\ the noise-free case in Fig.~\ref{fig:signal-mods}).
Model~{\MS} and all the overshoot models in the A and B sequences exhibit 
amplitudes that are larger than what is inferred for the Sun.
The models in the C sequence seem to span the range of solar amplitudes, 
with Model~{\OCtwo} providing the best fit.
Model~{\OP} has a smaller amplitude than is inferred for the Sun,
implying that the stratification in this model is even smoother
than the average spherically symmetric radial stratification in the Sun.

It is more problematic to draw inferences from a comparison of model
and solar values of the acoustic depth parameter $\btaud$, because 
the acoustic depth of the base of the convection zone is also strongly 
influenced by the sound-speed stratification in the rather uncertain
near-surface layers.
Specifically, the total acoustic size of the model and the acoustic 
location of the upper reflection boundary for each mode will have
an impact on the value found for this parameter.
Therefore its conversion into radial distance from the centre is not straightforward.
We recall the values of $\btaud$ obtained for Models {\MS} and {\MSpr}, shown in Fig.~\ref{fig:signal-mods}, with a shift in the fitted values of $\btaud$ (error free data) of about 25~s.
The only difference between these two models occurs at the surface, in order to bring the frequencies of {\MSpr} much closer to the solar frequencies.
Consequently this value is an indication on the accuracy we may expect on $\btaud$.

\begin{figure}
\centering
   \includegraphics[width=\hsize]{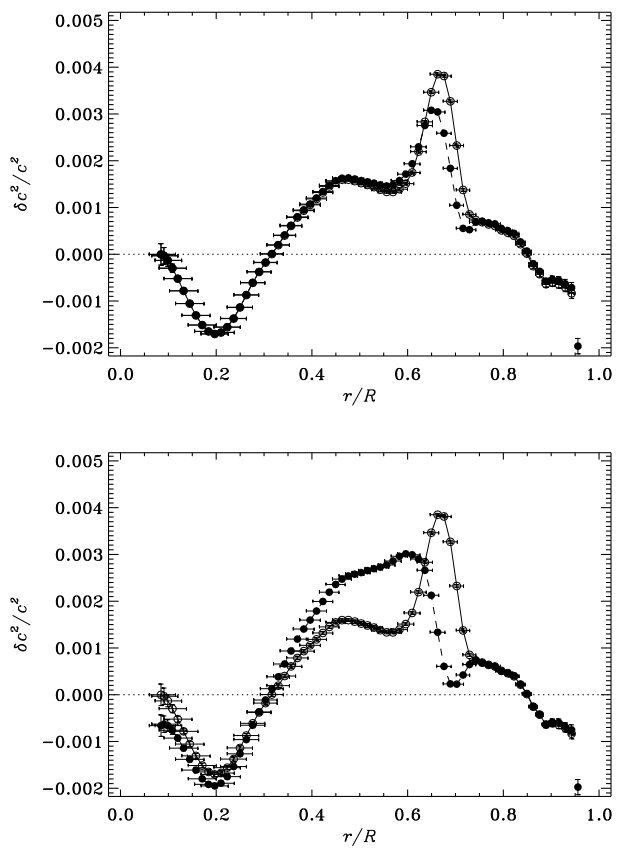}
\caption{
Relative differences $\delta c^2/c^2$, inferred by means of SOLA inversion,
between the Sun and three solar models, in the sense (Sun) -- (model).
The analysis used the so-called `Best Set' of observed frequencies of
\citet{Basu1997}.
The vertical bars show the $1\,\sigma$ errors in the differences, estimated 
from the errors in the observed frequencies, while the horizontal bars provide
a measure of the resolution of the inversion.
In both panels the open symbols and solid line shows results for Model~{\MS}.
The closed symbols and dashed line are for Model~{\OAtwo} (upper panel)
and Model~{\OCtwo} (lower panel).
}
\label{fig:inv}
\end{figure}

The fitting of the signal in the frequencies stills suffers from some shortcomings, namely the incomplete representation of the dependence of the amplitude and argument of the signal on mode frequency and degree. 
On the other hand, the approach we have considered here uses of a large number of modes (over 240), by including data up to mode degree of 20.
Also we isolate the signal directly in the frequencies avoiding noise amplification that occurs when the signal is isolated in frequency differences.
Both aspects are unique to our approach allowing in the case of the Sun a direct and very precise study of the base of the solar convection zone.
However, it is evidently also of great interest to investigate the extent to 
which information about the base of convective envelopes can be obtained from
just low-degree data, as available in the foreseeable future from observations
of other stars \citep[see][]{Monteiro:etal:2000}.

In this investigation we are using seismic data to look 
at only one aspect of the solar structure, namely the sharpness and shape of the transition in stratification near the base of the convection zone. 
The seismic data contain much additional information about the solar interior,
notably about the run of adiabatic sound speed with depth. 
The sound-speed differences between the Sun and three of the models considered here -- Models~{\MS}, {\OAtwo} and {\OCtwo} -- inferred from
helioseismic inversion of the data are illustrated in Fig.~\ref{fig:inv}.
The analysis was carried out using the so-called SOLA technique
\citep[see][for details on the implementation]{Rabell1999}, following 
\citet{Christ2007} and using the combined `Best Set' of
frequencies \citep{Basu1997} extending to $l = 99$.
Although there are statistically significant differences
in sound speed between the Sun and Model~{\MS}, these differences are small (0.4\% in sound-speed squared, {\ie}
0.2\% in sound speed). The differences between the illustrated models ({\OAtwo} and {\OCtwo}) and the Sun are similarly
small. In fact, although the models in this paper span a large range of behaviours in terms of the
stratification near the base of the convection zone, 
we stress that they are all close to the Sun in terms of
their small absolute differences in sound speed.

We remark that the details of the sound-speed gradient arising from overshoot
in some cases
may have substantial effects on the seismically inferred sound-speed 
differences between the Sun and the model.
In Model~{\OCtwo}, illustrated in the lower panel of Fig.~\ref{fig:inv} and
in Fig.~\ref{fig:nabcsq}, 
the region where $\nablacsq$ is larger than in Model~{\MS}
leads to an increase in the sound speed just below the convection zone which
largely eliminates the bump in $\delta c^2$ found with Model~{\MS}.
However, if this overshoot region were to be shifted to greater depth, thus
eliminating the region of sub-adiabatic gradient in the lower part of the 
convection zone, the effect on the sound speed would be amplified to such an 
extent as to lead to a strong {\it negative} sound-speed difference between
the Sun and the model. (To avoid this for the very smooth Model~{\OP} we
had to invoke a small opacity reduction to shift the location of the
boundary of the convection zone.)
Thus it remains important to test the models both based on the properties
of the signal associated with the acoustic glitch and based on the
inferred sound-speed differences.

\subsection{Implications for the stratification at the transition}

As found by \cite{ChristensenDalsgaard:etal:1995} standard models of the Sun either without or with the classical representation of overshoot,
based on a non-local mixing-length formulation,
are marginally inconsistent with the seismic data.
In Fig.~\ref{fig:signal-obs}
both the amplitude for some models and for the solar data is shown.
The reference Model~{\MS} is outside the error bar of the data as well as any of the overshoot models calculated in the classical way.
Even with a radiatively stratified overshoot layer the amplitude is found to be above the value found for the Sun.
This is a clear indication that the transition at the base of the envelope requires a representation that is smooth over length scales of the order of the wavelength of the modes.

This clear preference is a strong indication in favour of stratifications that include a sub-adiabatic layer within the proper convection zone, as a necessary condition to obtain sufficiently smooth transitions that will be consistent with the solar data.
Any model that includes overshoot will need to be extended from within the convection zone, replacing the commonly adopted procedure of adding an overshoot layer to the convective envelope as defined by the mixing-length theory (or equivalent).

The strong helioseismic preference for rather smooth profiles puts very strong 
constraints on many of the models we discussed in Sect.~\ref{sec:intro}. In the following
discussion we shall try to link the smoothness of the transition in the
overshoot region to the underlying physical processes.

A common physical process relevant in most overshoot models is buoyancy braking,
which decelerates downflow plumes once a positive temperature perturbation has
been built up after travelling downward for a sufficient distance in a subadiabatic
stratification. Buoyancy breaking establishes a direct connection between the
kinetic energy of downflows at the base of the convection zone and thermal
properties of the overshoot region beneath. This leads to a relation
between the Mach number $M_a$ required at the base of the convection zone, the
subadiabaticity as well as the depth of the overshoot region. Assuming for
simplicity a constant value for the superadiabaticity $\delta=\nabla-\nabla_{\rm ad}$, 
we can estimate temperature and density perturbations as function of the overshoot
depth, $z$, as follows:
\begin{equation}
\frac{T^{\prime}}{T}\sim \frac{\varrho^{\prime}}{\varrho}
\sim\delta\frac{z}{H_p} \label{thermal}\; .
\end{equation}
Buoyancy braking gives a relation between the kinetic energy of the plume at
the base of the convection zone and the buoyancy work in the overshoot region:
\begin{equation}
\frac{1}{2}\varrho v^2 \sim \frac{1}{2}g\varrho^{\prime} z=\frac{1}{2} \varrho\delta g H_p
\left(\frac{z}{H_p}\right)^2 \; ,
\end{equation}
where $g$ is the gravitational acceleration.
Combining both equations leads to (neglecting factors of order unity):
\begin{equation}
 M_a^2\sim \delta \left(\frac{z}{H_p}\right)^2\label{Ma-delta}\; .
\end{equation}
In the traditional non-local mixing-length overshoot models we have 
$z\simeq 0.2 H_p$ and $\delta\simeq 10^{-6}$ leading to 
$M_a\simeq 2\times 10^{-4}$, which corresponds to a convective velocity of about $40 \,{\rm m \, s^{-1}}$,
consistent with mixing-length models assuming a filling factor of about $0.1$.
Having a smooth transition of the overshoot from nearly adiabatic
to radiative temperature gradients requires that the average values for $\delta$ are of the order of $0.05$.
In that case our estimate requires $M_a\simeq 0.05$ corresponding to a convective velocity 
of about $10 \, {\rm km \, s^{-1}}$ at the base of the convection zone.
The latter is (if at all) only possible if the filling factor for downflow plumes is very small.
This is a inevitable consequence of buoyancy braking, which applies to overshoot models in a very 
general sense, whether numerical 3D simulation or any other simplified treatment.
The key for the argumentation above is the fact that a smooth overshoot profile implies
automatically substantial deviations from adiabaticity in an extended region that is
convectively mixed.
Many of the numerical overshoot simulations to date show smooth transitions. However,
the latter are to a large degree related to numerical setups that have either a reduced
stiffness in the radiative interior or have large energy fluxes (to reduce thermal relaxation
time scales) that result in much larger Mach numbers and allow consequently for smooth
overshoot profiles with substantial deviations from adiabaticity. \cite{Kapyla2007} presented a
series of numerical experiments that show the return to more adiabatic step-like overshoot
profiles with reduction of the overall energy flux. Extrapolating these results to the solar
energy flux seems to make a step-like quasi-adiabatic overshoot almost inevitable. The investigation
of \cite{Rempel:2004} aimed at understanding the physical parameters under which a
smooth transition can be expected. The two necessary conditions found were 
that a) the 
downflow filling factor is small and b) there is a continuous distribution of downflow
strength. While the latter is a natural consequence of convection, the former is very 
controversial and not supported by any numerical simulations we know about (the required
downflow filling factors would be smaller than $10^{-4}$). 

Overall it appears that within the framework of models discussed above it is almost impossible
to obtain sufficiently smooth profiles that agree with the helioseismic constraints.
This difficulty is strongly related to the effective role of buoyancy braking in the overshoot
region. Only if we relax this assumption is it possible to obtain smooth 
overshoot profiles for moderate Mach numbers. It has been suggested by \citet{Petrovay:Marik:1995} 
that smooth overshoot profiles can be easily obtained once the assumption of a strong 
correlation between vertical velocity and temperature fluctuation is dropped. They suggested that in
the extreme case of vanishing correlation the penetration is only limited by turbulent dissipation, 
with no impact on  the thermal stratification. For a smoothly diminishing correlation a smooth 
transition of all overshoot quantities results.   

\citet{Xiong2001}, in fact, presented a model that leads
to such very smooth transitions, by using a non-local convection model that is based on a moment 
approach computing auto and cross-correlations of turbulent velocity and temperature
\citep{Xiong:1989}. This model predicts a strong decorrelation of velocity
and temperature fluctuation in the lower overshoot region of the Sun,
while the authors find 
a strong anticorrelation in the upper overshoot region above the photosphere.
The difference between the
overshoot regions is attributed to the values of the P\'eclet number, which are much smaller than unity
in the photosphere and much larger in overshoot region at the base of the convection zone.
The overall outcome is a temperature profile that is very similar to the best match we found in this
investigation. Their overshoot region extends about $0.6 H_p$ beneath the convection zone and
shows already a significantly subadiabatic stratification within the lower parts of the convection
zone.  Recently \citet{Marik:Petrovay:2002} used a different non-local model based on the
formulation of \citet{Canuto:Dubovikov:1997} and \citet{Canuto:Dubovikov:1998} and found overshoot
of only $0.06 H_p$. While still much smoother than the sharp profiles of non-local mixing-length theory,
this model provided again a temperature profile that stayed closed to adiabatic for significant extent
of the overshoot depth.
{\rbf This class of models involves a number of assumptions,
including the treatment of closure, and parameters generally determined
from laboratory experiments or atmospheric or oceanographic measurements;
thus a thorough study of their applicability under stellar conditions
is indicated. In addition, a further investigation is required of
the differences between the
\citet{Xiong2001} and the \citet{Canuto:Dubovikov:1997} approaches.}

Another possibility could be a spatial or temporal inhomogeneity of overshoot leading
to an average profile looking more smooth; however as estimated below this
possibility seems unreasonable, too:

Assume that we have a large filling factor $\sim 0.1$ so that the overshoot
profile has a very sharp transition. Intermittent downflows are able to
perturb the sharp boundary between the almost adiabatic overshoot and the
underlying strongly subadiabatic radiative zone. The consequence are
buoyancy oscillations with the Brunt V\"asail\"a frequency
$\omega_{\rm BV}=\sqrt{g\vert\delta\vert/H_p}$.
Due to the sharp transition
toward the radiation zone, $\omega_{\rm BV}$ would be given by the radiative
values just beneath the overshoot region. With a value of
${\rm d}\nabla_{\rm rad}/{\rm d}r=10^{-8} \,{\rm m^{-1}}$
at the base of the convection zone, 
overshoot with a depth $d_{\rm os}$ leads to a value of
$\delta=-({\rm d}\nabla_{\rm rad}/{\rm d}r)\,d_{\rm os}$. Using
$d_{\rm os}=0.1\,H_p$, $g=500\,{\rm m}\,{\rm s}^{-2}$,
$H_p=60\,{\rm Mm}$ yields 
$\omega_{\rm BV}\sim 7\times 10^{-4}\,{\rm s}^{-1}$. Amplitude $d$ and
velocity $v$ of the buoyancy oscillations at the overshoot radiative interior
interface are related by $v=\omega_{\rm BV}d$, where $v$ should be of the order
of the convective motions perturbing the interface. Using a value of
$v=100\,{\rm m}\,{\rm s}^{-1}$ yields $d=140\,{\rm km}$, which is comparable to
the thickness of the transition expected from the overshoot model in the first place.
A larger amplitude requires a larger $v$, which can be only obtained
by having faster downflows, meaning smaller filling factor (so nothing is
gained here). We note that the above discussion implicitly assumed again a strong
correlation between temperature and velocity fluctuations.

It has been also speculated that the depth of overshoot depends on latitude, due
to the changing influence of rotation on convective motions. The study of
\cite{Brummell:etal:2002} found the deepest penetration at pole and equator and a 
minimum at around $30^\circ$ latitude with about $60\%$ of the maximum overshoot depth.
This result was obtained by placing small Cartesian simulation domains (with periodic
boundaries in the horizontal direction) at different latitude positions. Having
a variation of overshoot depth in a global model adds additional complexity
to the problem, since regions with different vertical temperature structures
have to be in horizontal force balance. These temperature differences can be
estimated from Eq. (\ref{thermal}) and can be quite substantial. Restricting
$T^{\prime}$ to $100\,{\rm K}$ allows with $\delta\sim 0.1$,
$T\sim 2\times 10^6\,{\rm K}$
and $H_p=50\,{\rm Mm}$ for only a displacement of $25\, {\rm km}$.
A global force balance
in latitude would require either zonal flows or magnetic forces of substantial 
strength -- at least the zonal flows should leave observable helioseismic signatures
given the fact that only about $10 \, {\rm K}$ temperature difference are required in the
convection zone to balance the deviations of differential rotation from the
Taylor-Proudman state. Without these flows or magnetic field such a configuration 
cannot be in a hydrostatic balance and would return to spherical symmetry on a time 
scale given by $\omega_{\rm BV}$.



\section{Conclusions}

We have considered a variety of models with different stratification near the base of the convection zone, using a parameterization of the stratification that is motivated by models that have a spectrum of overshooting plumes with a range of depths of penetration.
{}From the range of models we have considered, we find that a relatively smooth model (Model~{\OCtwo}) fits the helioseismic data better than a standard model without convective overshoot, and better also than simple convective overshoot models which possess a more-or-less sharp transition to subadiabatic stratification beneath an overshoot region which is a nearly adiabatic extension of the convection zone.
Thus a model characterized by an overshoot layer extending for $\ell_{\rm ov}/H_p \simeq 0.37$ is compatible with the solar data, corresponding to an additional fully mixed zone of about 3\% in radius.
Of course it is desirable that solar models be built whose stratification is determined self-consistently from the physics underlying the model, 
but the parameterization used here may be a reasonable way of approximating such stratification in the context of simple spherically symmetric models. 

Our investigation method is very similar to that developed by \cite{Monteiro:etal:1994}, though with some differences due to the fact that here we determine the phase $\phi_0$ independent of our fitting to the oscillatory signal in the frequencies, and therefore we fit amplitude parameters $a_1$ and $a_2$ separately.
It should though be noted that we find that the fitted values of $a_1$ and $a_2$ are not independent, probably because the frequency range of the modes used in the fitting is rather small and insufficient fully to separate the frequency dependent terms of the amplitude, and so their separate values cannot be relied upon.
Indeed we find that noise in model data tends to cause the value of $a_2$ to be significantly underestimated, relative to its fitted value for noise-free data.
We find, however, that the combination amplitude $\A25$ is a more robust indication of the sharpness of the transition in stratification near the convection-zone base.


Our selection of models demonstrates that the signal analysed here is
sensitive to acoustic glitches that would barely be noticed through
helioseismic inversion which provides averages over such features.
On the other hand, our analysis is by design insensitive to smooth
differences between the Sun and the model, including very substantial
differences in the sound speed which would be obvious from an inverse analysis.
Thus these two techniques are obviously complementary in their ability
to characterize the properties of the solar interior.

It may be remarked that we have used models similar to Model~{\MS}, which incorporates the ``old' solar element abundances of \citet{Greves1993},
rather than the ``new'' abundances suggested by the work of Asplund and collaborators \citep[e.g.,][]{Asplun2009}.
The new abundances are known to modify the opacities in the solar interior in such a way as to produce models that are strongly in disagreement 
with the stratification inferred from helioseismology 
in the radiative interior, the largest discrepancy being in the 
vicinity of the base of the convection zone.
Since our aim in this work is to investigate the subtle effects of overshoot in this region, it is prudent to start with models that are broadly consistent with the known stratification of the Sun's interior, rather than models based on the newer abundances that are further away from the Sun's stratification. 


Although the present analysis makes use of the full range of modes that are
sensitive to the base of the convection zone, similar analyses are possible
given just the low-degree modes available in observations of distant
stars in the foreseeable future \citep[e.g.,][]{Monteiro:etal:2000,
Ballot2004, Piau:etal:2005}.
This is potentially an important complement to the more detailed solar
data, particularly given the very long timeseries and hence high frequency
precision that will be obtained with the {\it Kepler} mission
\citep[e.g.,][]{Gillil2010} as well as from planned
dedicated ground-based facilities.


{}From this investigation we can conclude that i) overshoot is necessary to
improve the agreement between models and helioseismic constraints, ii)
the required overshoot profiles are outside the realm of the classic
``ballistic" overshoot models and iii) the lower part of the convection
zone is likely substantially subadiabatic. We cannot measure the latter 
directly, but our investigation indicates that the required level of smoothness
cannot be achieved without iii). Currently only non-local convection models
that are based on auto- and cross-correlations of velocity and temperature 
perturbations are capable of providing overshoot profiles with the desired degree
of smoothness in the transition. However, these theories have hidden parameters
and rely on closure models. To our current knowledge there has not yet been
a thorough study on how robust the properties of overshoot (under the 
conditions found at the base of the solar convection zone) are within
the framework of these models. We hope that our investigation motivates more
research in that direction.
The indication that the lower part of the convection zone could be 
substantially subadiabatic (with values of 
$\vert \nabla-\nabla_{\rm ad}\vert>10^{-3}$) has profound consequences for
the storage and stability of magnetic field and the overall role that the
lower convection zone (not just the overshoot region) might play in the solar 
magnetic cycle.   

\section*{Acknowledgments}

We thank the referee for constructive and helpful comments on an
earlier version of the manuscript, which have improved the presentation.
MJM was supported in part through project
{\small PTDC/CTE-AST/098754/2008}, funded by FCT/MCTES Portugal and by the European programme FEDER.
{\rbf This work was supported by the European Helio- and Asteroseismology Network (HELAS),}
a major international collaboration funded by the European Commission's Sixth Framework Programme.


\appendix
\section[]{Determining the phase of solar eigenfunctions}

\subsection[]{The eigenfunction phase}
\label{sec:eigphase}
The phase $\phi(\omega,l)$ in Eq.~(\ref{eq:eigphase}) is obviously
closely related to the eigenfunctions of the oscillations and
hence can be determined from fits to those eigenfunctions.
This was discussed in detail by \cite{CDPH92} 
\citep[see also][]{Roxbur1996},
in the context of the \cite{Duvall1982} law.
According to this, the frequencies satisfy
\begin{equation}
\int_{r_{\rm t}}^R \left( 1 - {L^2 c^2 \over \omega^2 r^2} \right)^{1/2}
{\dd r \over c} = {\pi[n + \alpha(\omega, l)] \over \omega} \; ,
\label{eq:duvall}
\end{equation}
where $L^2 = l(l+1)$.
This is obtained from Eq.~(\ref{eq:eigenf}) by applying the
appropriate inner boundary conditions, with $\phi$ and $\alpha$
being related by
\begin{equation}
\phi = - \left(\alpha + {1 \over 4} \right) \pi \; .
\label{eq:duvphase}
\end{equation}
For simplicity we neglect the higher-order terms in the near-surface 
behaviour and assume that $\alpha$, and hence $\phi$, are functions
of $\omega$ alone.
The quantity $\phi_0$ entering into the fitting formula 
is the intercept at $\omega = 0$ in a linear fit to $\phi$ as a
function of $\omega$ and hence can be determined from
a similar fit to $\alpha$.

\citet{CDPH92} analysed the dependence of $\alpha$ 
on the properties of solar models
by fitting Eqs (\ref{eq:eigenf}) and (\ref{eq:eigphase}) to computed
solutions of the equations of radial oscillations in the models,
assuming only the surface boundary conditions and hence determining
$\alpha$ as a continuous function of $\omega$
\citep[a similar procedure has been used by][]{Roxbur1996}.
Here we consider instead full eigenfunctions of the model, over a range
of degrees, to approximate more closely the actual fits.
As a simplification, we use only relatively low-degree modes and apply
the fit rather close to the stellar surface. 
Then Eqs (\ref{eq:eigenf}) and (\ref{eq:eigphase}) can be approximated by 
\begin{equation}
\xi_r = A (\rho c)^{-1/2} r^{-1} \cos(\omega \tau + \phi) \; ,
\end{equation}
in terms of the acoustic depth $\tau$ (cf.\ Eq.~\ref{eq:tau}).
We therefore fit $A \cos(\omega \tau + \phi)$ to $E = (\rho c)^{1/2} r \xi_r$
in a least-squares sense, over a suitable interval $[\tau_1, \tau_2]$ in $\tau$.

\begin{figure}
   \centering
   \includegraphics[width=\hsize]{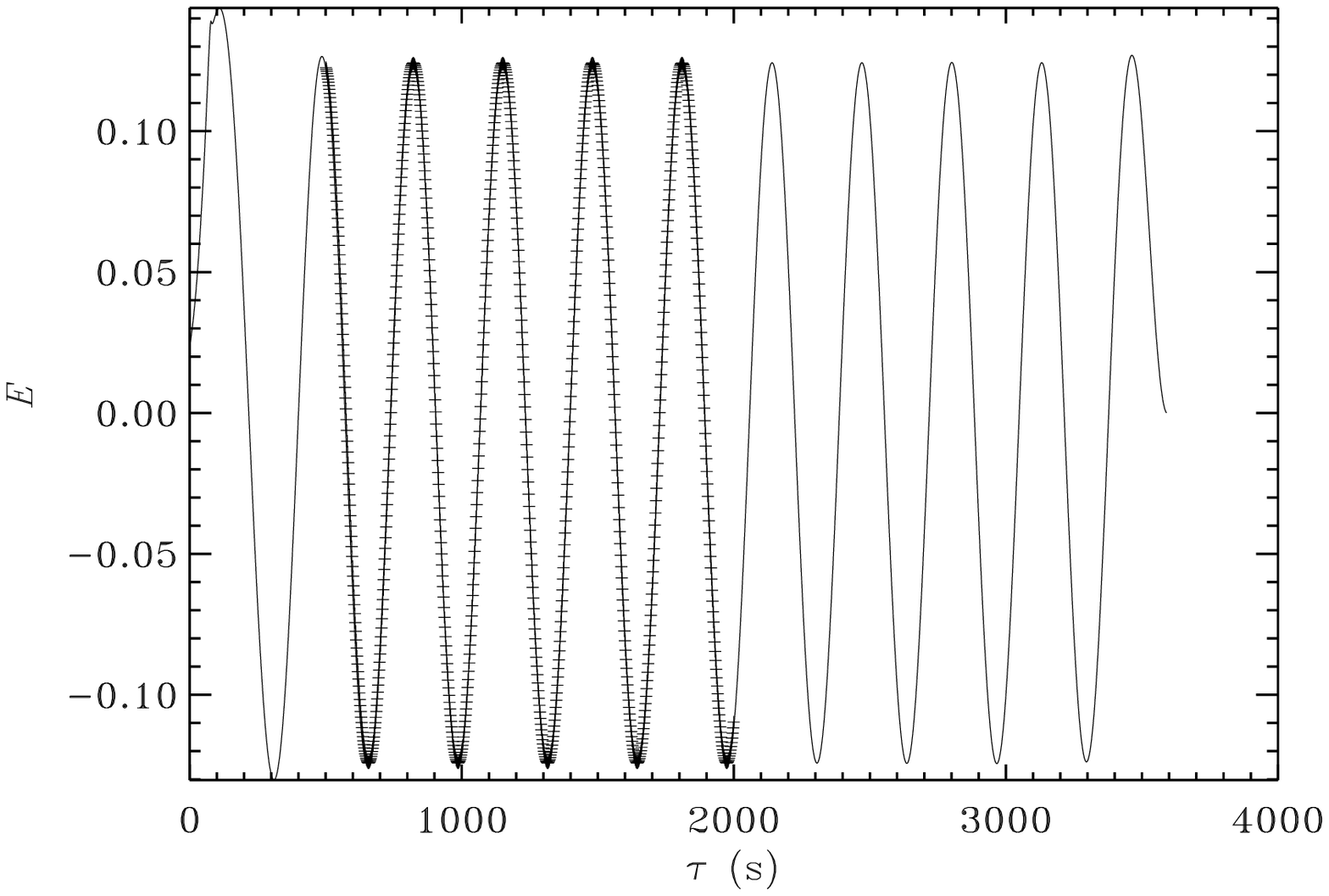}
   \caption{
Scaled eigenfunction $E = (\rho c)^{1/2} r \xi_r$,
for the mode with $l = 0$, $n = 21$ in Model~{\MS},
as a function of acoustic depth $\tau$.
The crosses show the results of fitting the asymptotic eigenfunction.
   \label{fig:eigfit}}
   \end{figure}

As an example we consider Model~{\MS} \citep{Christ1996}.
Figure \ref{fig:eigfit} shows an example of the fit to an eigenfunction,
for the mode with $l = 0$, $n = 21$.
The fit was made in the interval between $500 {\,\rm s}$ and $2000 {\,\rm s}$
 in $\tau$.
As illustrated by the crosses, the fit is excellent.

\begin{figure}
   \centering
   \includegraphics[width=\hsize]{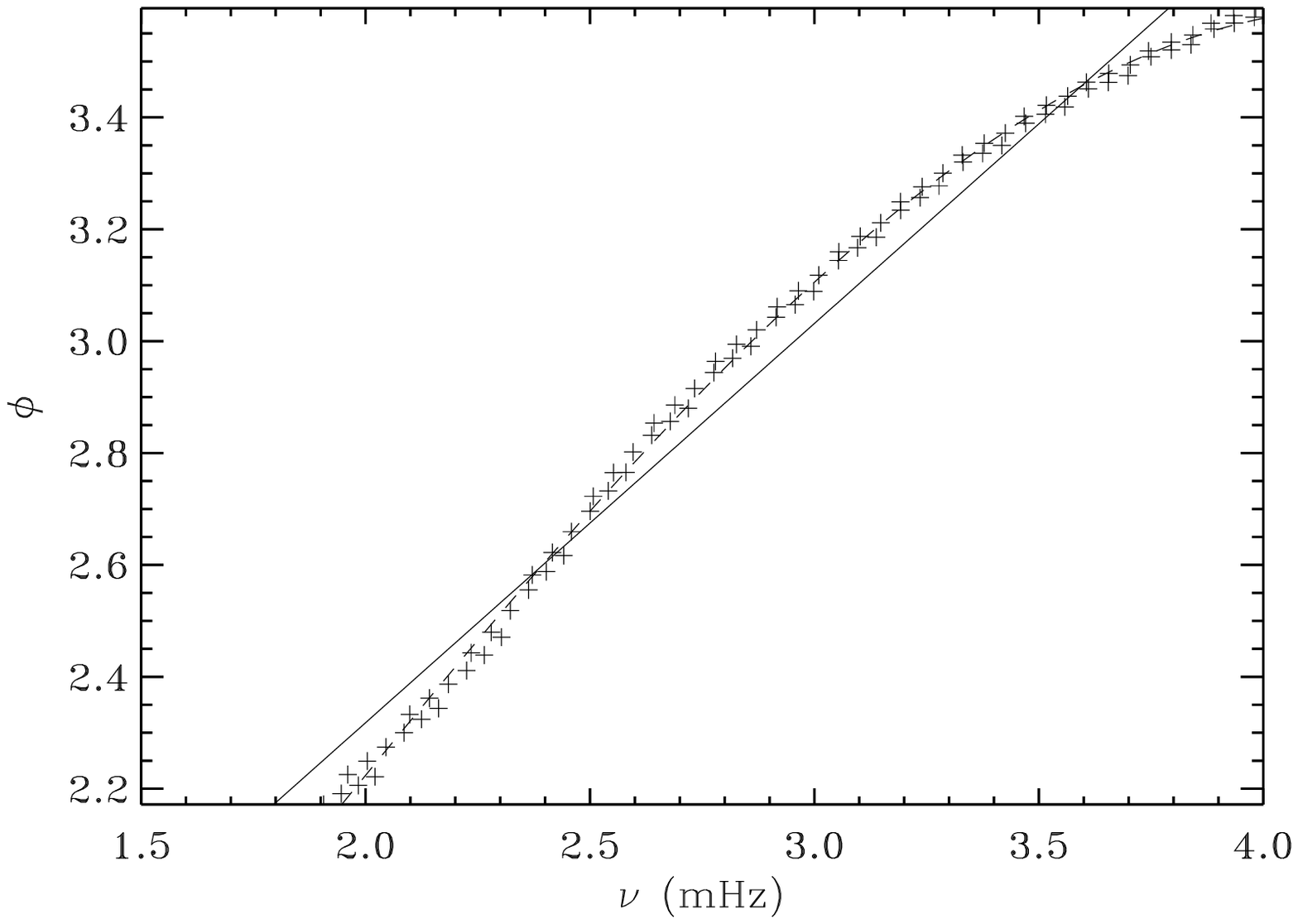}
   \caption{
Fitted phases for modes between 5 and 10 in degree and 1.9 and 4 mHz in
cyclic frequency $\nu = \omega/2 \pi$, in Model~{\MS}.
The solid and (barely visible) dashed curves show linear and 
cubic least-squares fits to the points, respectively.
   \label{fig:sphase}}
   \end{figure}

To determine $\phi(\omega)$ we consider modes in the interval
$5 \le l \le 10$; also, in accordance with the mode sets used in
the fit for the base of the convection zone, we consider frequencies
between 1.9 and 4 mHz.
The phases resulting from the fit are shown in Fig. \ref{fig:sphase},
confirming that the scatter at given frequency is modest.
Also shown is a least-squares fit of a straight line which yields
$\phi_0 = 0.890$,
and a cubic fit which follows the computed points
very closely but yields an intercept at $\omega = 0$ of 0.503.

\begin{figure}
   \centering
   \includegraphics[width=\hsize]{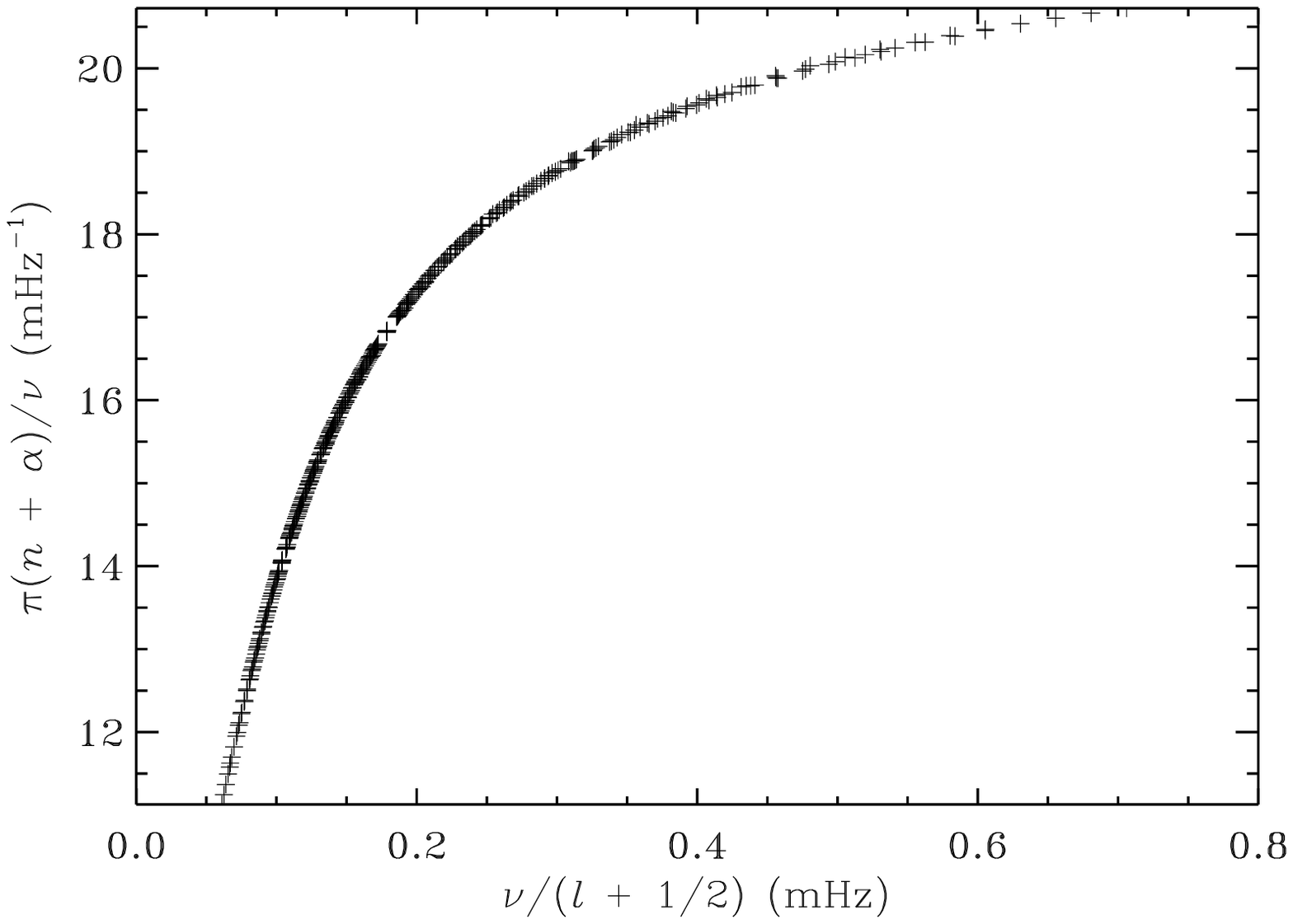}
   \caption{
Duvall plot for the modes between 5 and 100 in degree and 1.9 and 4 mHz in
frequency, in Model~{\MS}, 
determining $\alpha$ from the cubic fit to the phase in
Fig. \ref{fig:sphase}.
   \label{fig:sduvall2}}
   \end{figure}

The Duvall relation resulting from $\phi$, and hence $\alpha$,
determined in this manner is illustrated in Fig.~\ref{fig:sduvall2},
using $\alpha(\omega)$ as obtained from the cubic fit in 
Fig. \ref{fig:sphase}.
Here all modes with $5 \le l \le 100$ and frequency between 1.9 and 4 mHz
are included.
The match to the Duvall law is very good,
strongly indicating that $\phi$ as determined from the low-degree modes
can be applied over a broad range of degrees.

\subsection[]{Differential phase determination}

While the eigenfunction fit provides the most appropriate 
phase $\phi(\omega)$ for the models, and hence presumably an appropriate
estimate of $\phi_0$, this is obviously not possible for the observations.
However, the results above suggest that an estimate of $\phi$ and $\phi_0$
can be obtained from a fit of the Duvall relation.
One possibility would be to determine the function $\alpha(\omega)$ that
most successfully collapses the points in the Duvall plot to a curve and
to obtain $\phi(\omega)$ from that.
A possibly simpler approach is to use the differential form of the
Duvall law \citep{CDGPH88} to determine a correction to $\phi_0$,
relative to a suitable
reference model.
By linearizing Eq.~(\ref{eq:duvall}) in small changes $\delta c$ and
$\delta \alpha$ we obtain
\begin{equation}
S {\delta \omega \over \omega} = \CH_1(\omega/L) + \CH_2(\omega) \; ,
\label{eq:difduvall}
\end{equation}
where
\begin{equation}
S =
\int_{r_{\rm t}}^R \left( 1 - {L^2 c^2 \over \omega^2 r^2} \right)^{-1/2}
{\dd r \over c}  - \pi {\dd \alpha \over \dd \omega} \; ,
\label{eq:s}
\end{equation}
\begin{equation}
\CH_1(w) =
\int_{r_{\rm t}}^R \left( 1 - {c^2 \over w^2 r^2} \right)^{-1/2}
{\dd r \over c} \; ,
\label{eq:h1}
\end{equation}
and
\begin{equation}
\CH_2 (\omega) = {\pi \over \omega} \delta \alpha(\omega) \; .
\label{eq:h2}
\end{equation}
As discussed by \cite{CDGT89} the functions
$\CH_1$ and $\CH_2$ can be approximated by a double-spline fit to
the scaled frequency differences, to obtain $\bar \CH_1$ and $\bar \CH_2$.
Given $\bar \CH_2$, Eqs (\ref{eq:duvphase}) and (\ref{eq:h2}) show that
the difference in the phase can be obtained as
\begin{equation}
\delta \phi = - \pi \delta \alpha \simeq - \omega \bar \CH_2 \; .
\label{eq:difphase}
\end{equation}

It should be noticed that a fit of the form given in
Eq.~(\ref{eq:difduvall}) can only determine the functions $\CH_1$ and
$\CH_2$ to within a constant.
Consequently, $\delta \phi$ as determined from Eq.~(\ref{eq:difphase})
is only determined to within a constant multiple of $\omega$.
Obviously, this has no effect on the inferred value of $\delta \phi_0$.


\begin{figure}
   \centering
   \includegraphics[width=\hsize]{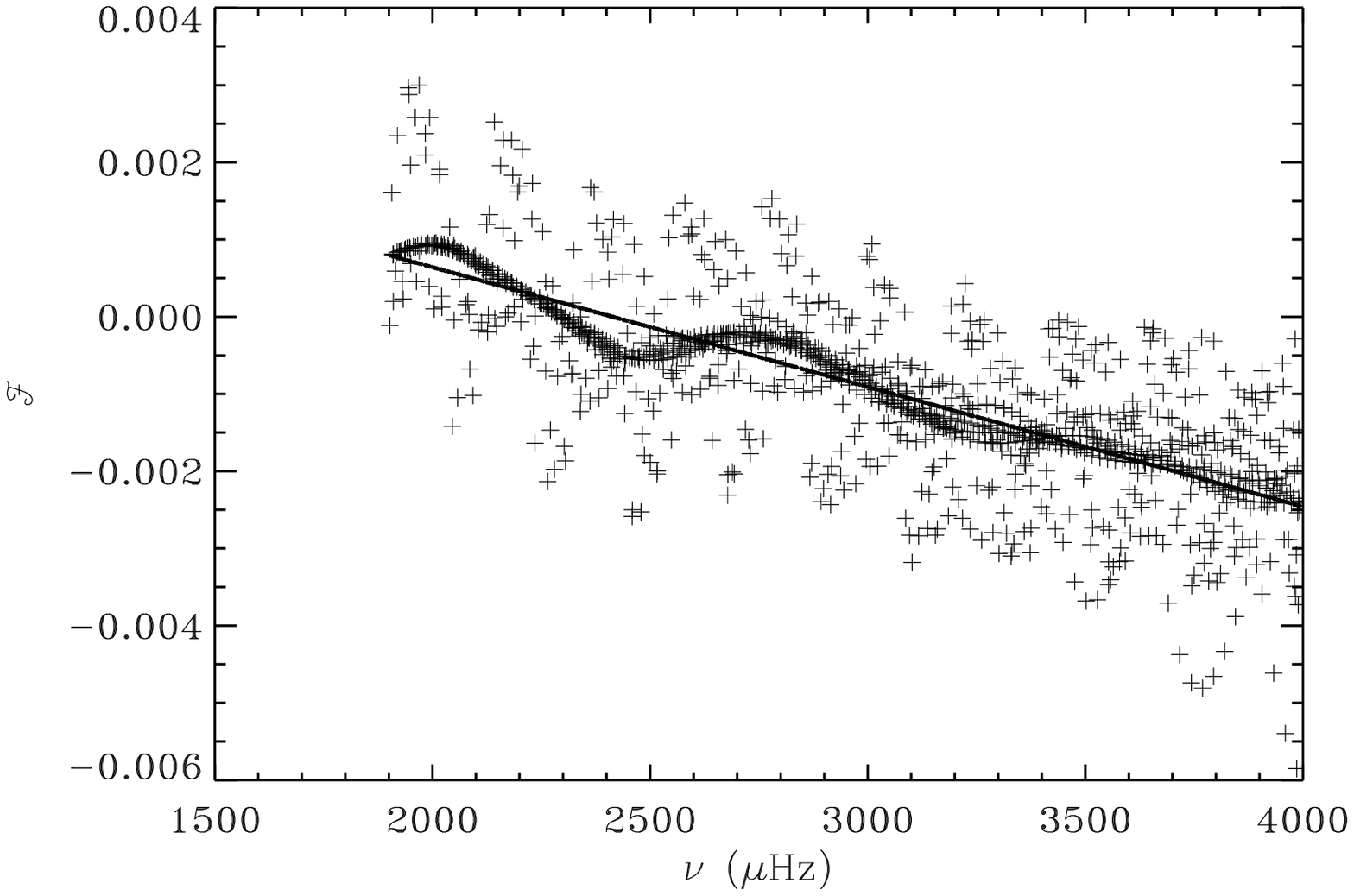}
   \caption{
The predominantly frequency-dependent residual difference 
$\CF$ (cf.\ Eq. \ref{eq:resdif})
resulting from a fit to frequency differences
between Model~{\OCtwo} and Model~{\MS}, using Eq.~(\ref{eq:difphase}).
The straight line shows a linear least-squares fit to the results,
resulting in a determination of $\phi_0$.}
   \label{fig:difphase82}
   \end{figure}

\begin{table}
   \caption{Phases and phase differences resulting from eigenfunction and
differential Duvall fits;
$\delta \phi_0^{(1)}$ is the difference relative to Model~{\MS}
computed directly from $\phi_0$ {\rbf for the two models},
while $\delta \phi_0^{(2)}$ is the corresponding difference
inferred from the differential asymptotic Duvall fit (Eq.~\ref{eq:difphase}).
The solar result of the differential asymptotic analysis used
a 72-d MDI dataset starting in April 1996. 
}
   \label{table:a1}      
   \vskip 3pt
   \centering
\begin{tabular}{l l c r r}        
\hline
\noalign{\vskip 3pt}
Type & ID & $\phi_0$ & $\delta \phi_0^{(1)}$ & $\delta \phi_0^{(2)}$ \\
\hline
\noalign{\vskip 3pt}
Model \\
& {\MS}     & 0.8904 & -- & --  \\
& {\OCtwo} & 0.8944 & 0.0040 & 0.0037  \\
& {\OP} & 0.8954 & 0.0050  & 0.0047 \\
& {\MSpr} & 0.5485 & $-0.3419$ & $-0.3257$  \\
 \hline
 Sun &  & -- & -- & $-0.3181$  \\
\hline
\end{tabular}
\end{table}

We have tested this procedure by applying it to a few of the models
discussed in Section \ref{sec:models}.
Here the frequency range was, as usual, between 1.9 and 4 mHz, and modes
of degree between 0 and 100 were included.
For simplicity, the term in $\dd \alpha / \dd \omega$ was neglected
in computing $S$ from Eq.~(\ref{eq:s}).
To determine $\delta \phi_0$ we carry out the fit to determine
the function $\bar \CH_1(\omega/L)$ and evaluate 
\begin{equation}
\label{eq:resdif}
\CF = - (S \delta \omega - \omega \bar \CH_1) \; ;
\end{equation}
we then carry out a linear least-squares fit to $\CF$, as a function
of frequency, resulting in $\delta \phi_0$.
The procedure is illustrated in Fig. \ref{fig:difphase82}, 
for Model~{\OCtwo}.
In Table \ref{table:a1}, the result is compared with the actual difference
in $\phi_0$, relative to Model~{\MS},
obtained by fitting the eigenfunctions in the two models;
the agreement is clearly quite good.
Table~\ref{table:a1} also lists the very similar results for Model~{\OP}.

\begin{figure}
   \centering
   \includegraphics[width=\hsize]{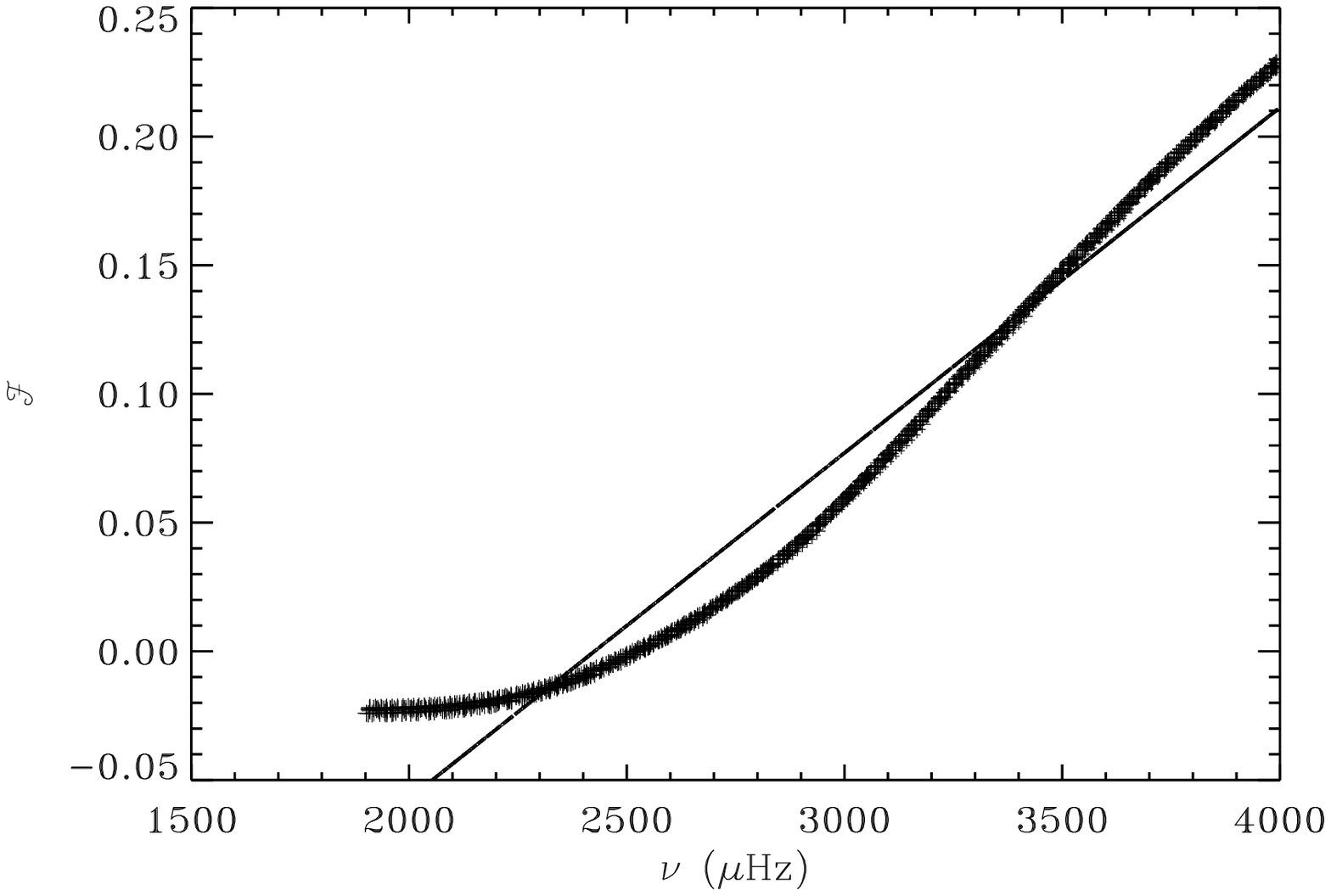}
   \caption{
The predominantly frequency-dependent residual difference 
$\CF$ resulting from a fit to frequency differences
between Model~{\MSpr} and Model~{\MS}, using Eq.~(\ref{eq:difphase}).
See caption to Fig.~\ref{fig:difphase82}.
   \label{fig:difphasegr1}}
   \end{figure}

As a more extreme example, we consider a model, Model~{\MSpr},
based on Model~{\MS} but modified to suppress the dominant, near-surface
part of the differences between the observed frequencies and those
of Model~{\MS} \citep[e.g.,][]{Christ1996}.
Specifically, $\Gamma_1$ was modified by multiplying it by
\begin{equation}
\label{eq:gammod}
1 - 0.4 \exp\left[ {(r {-} r_0)^2 \over \Delta r^2} \right] \; ,
\end{equation}
where $r_0 = (1 {-} 2.5{\times} 10^{{-}4}) R$
and $\Delta r = 1.5 {\times} 10^{{-}4} R$.
This was designed to emulate the effect of turbulent pressure on
the thermodynamics of the strongly superadiabatic part of the
convection zone (C.~S. Rosenthal, private communication).
The differences between the frequencies of this model
and those of Model~{\MS} essentially match the corresponding differences
for the observed frequencies and 
hence contain a large frequency-dependent component (after scaling).
This is reflected in the large frequency-dependent part of the
frequency differences and hence in the resulting $\CF$, 
illustrated in Fig. \ref{fig:difphasegr1}.
The value of $\delta \phi_0$ obtained from fitting to $\CF$
is listed in Table~\ref{table:a1},
together with the value resulting from the fits to the eigenfunctions.
Remarkably, even for this very substantial difference there
is reasonable agreement with the directly determined values of
$\phi_0$ for the models, indicating that the differential Duvall analysis
provides a robust method for estimating $\phi_0$, also for observed frequencies.

\begin{figure}
   \centering
   \includegraphics[width=\hsize]{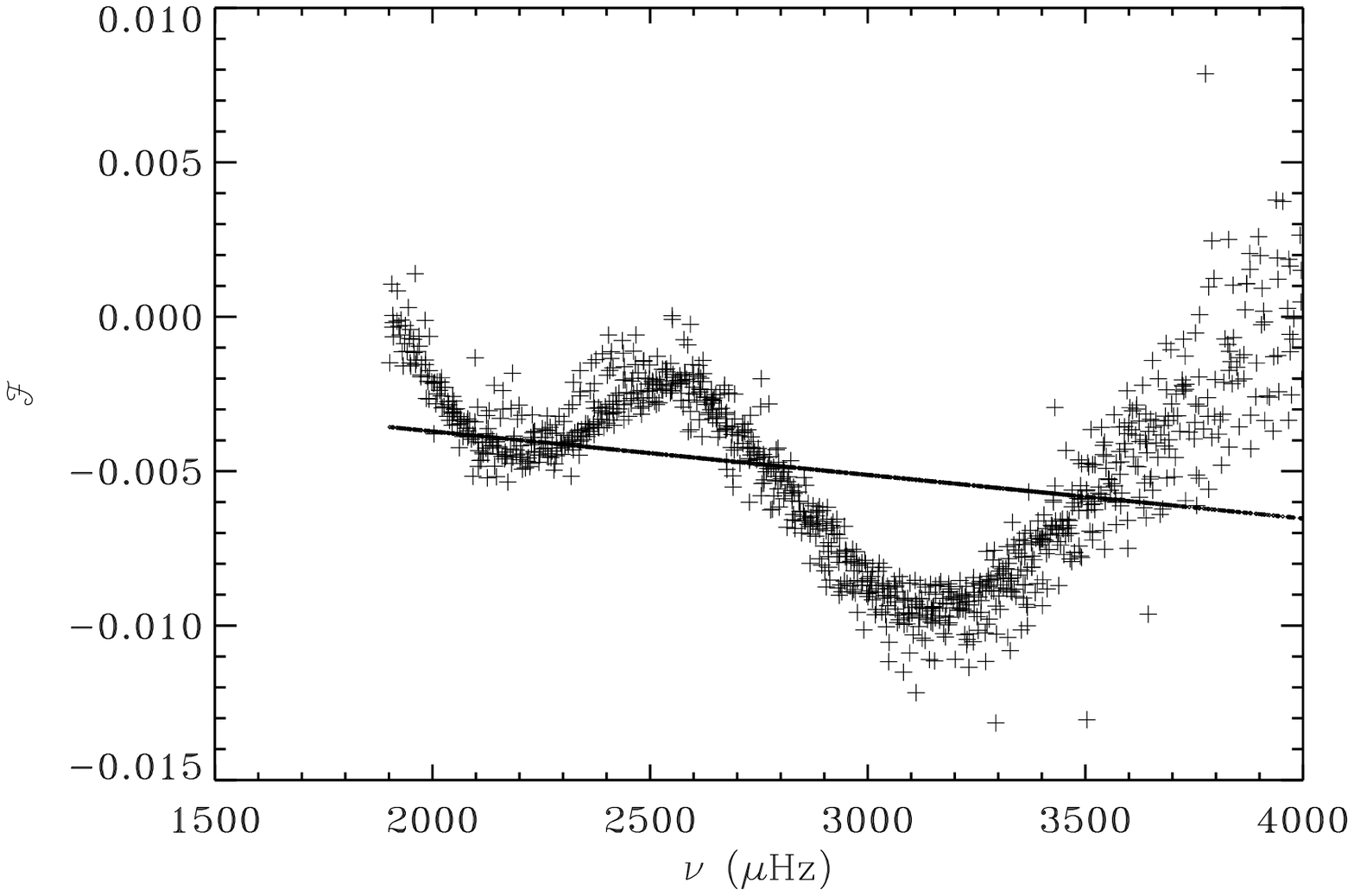}
   \caption{The predominantly frequency-dependent residual difference 
$\CF$ (cf.\ Eq. \ref{eq:resdif})
resulting from a fit to frequency differences
between MDI observations and Model~{\MSpr}, using Eq.~(\ref{eq:difphase}).
The observations were the 72-d set used in Table~\ref{table:a1}.
See caption to Fig.~\ref{fig:difphase82}.
   \label{fig:difphasemdigr1}}
   \end{figure}

We have also applied the differential Duvall analysis to a 72-d 
set of MDI observed frequencies.
As shown in Table~\ref{table:a1} this yields a $\delta \phi_0$ very similar
to that obtained for Model~{\MSpr}; 
this was to be expected, given that this model was fitted to the observed
frequencies.

In the fits of model frequencies to Eq.~(\ref{eq:signal}) we used the
value of $\phi_0$ resulting from a fit to the corresponding eigenfunctions.
In the analysis of the observed frequencies, we have used
Model~{\MSpr} as a reference;
in this way we hope to minimize any systematic effects that would 
result from the larger frequency differences relative to Model~{\MS}.
As an example, 
Fig.~\ref{fig:difphasemdigr1}
shows the phase-difference plot for one of the 72-d observational sets.
The linear fit gives $\delta \phi_0 = -  8.9 \times 10^{-4}$,
again indicating that Model~{\MSpr} provides a good fit
to the observations.
It is likely that the oscillatory component 
arises from a combination of the residual effects of
the acoustic glitch associated with the second helium ionization zone
\citep[e.g.,][]{Gough1990, Voront1991, Monteiro:Thompson:2005}
and a component of the near-surface difference which has not been
eliminated by the modification (\ref{eq:gammod}) to $\Gamma_1$.
The former effect would
indicate that the helium abundance (or the equation of state)
in the model (and hence in Model~{\MS}) is not in full agreement with that
of the Sun;
taken at face value the variation below $2500 \muHz$ suggests that the
envelope helium abundance of 0.245 in the model is too low by about 0.008.

\subsection{Solar-cycle variation}

\label{sec:solarcycle}
We have carried out the differential fit to determine the value
of $\delta \phi_0$ and hence $\phi_0$ for all the datasets involved in
the analysis.
The results for the 72-d sets were illustrated in Fig.~\ref{fig:obsdelphase}
which clearly reflects the variation in solar activity.
To understand the behaviour of the variation we compare
observed frequencies at an arbitrary phase
of the solar cycle with frequencies at solar minimum and recall that
the frequencies increase with solar activity, the change being
a steeply increasing function of frequency \citep[e.g.,][]{Libbre1990}.
We also note that $\CH_2$
essentially corresponds to the scaled relative frequency differences,
apart from the arbitrary constant in the separation into $\CH_1$
and $\CH_2$ in Eq.~(\ref{eq:difduvall}).
Thus $\CH_2$ increases with increasing frequency.
Choosing the constant such that $\CH_2$ is zero at the low end
of the frequency range considered,
the function $\CF$ is similarly zero at low frequency but with a negative 
slope. 
Consequently, the intercept $\delta \phi_0$ at zero frequency,
which is obviously unaffected by the choice of constant, is positive,
as observed in Fig.~\ref{fig:obsdelphase}, and increases with increasing
solar activity and hence increasing frequency.


\begin{figure}
   \centering
   \includegraphics[width=\hsize]{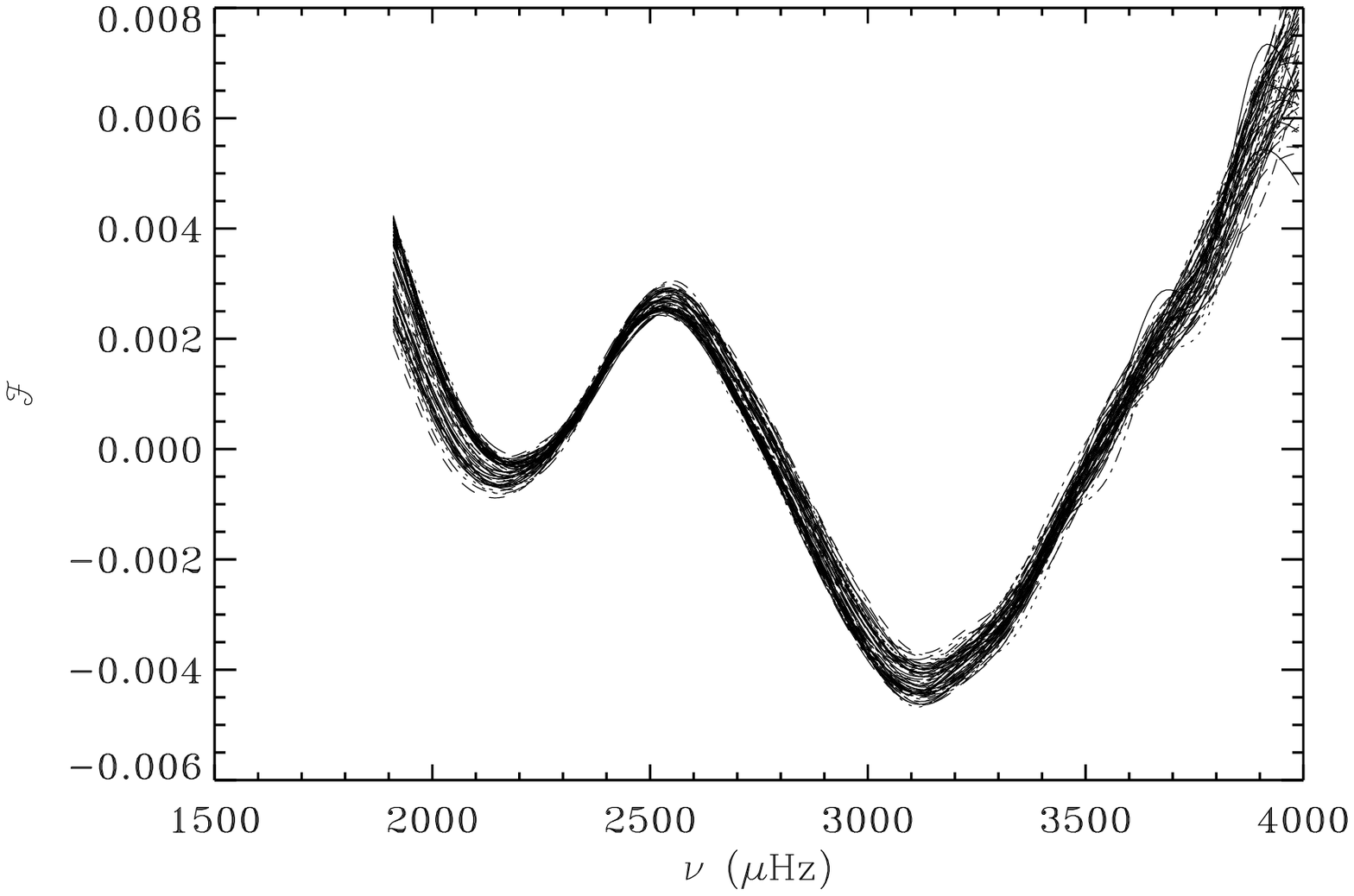}
   \caption{The functions $- \omega \CH_2$ resulting from
the spline fit (cf.\ Eq.~\ref{eq:difduvall})
to the frequency differences between all sets of 72-d observations
and Model~{\MSpr}, after subtraction of the linear fits to the
corresponding $\CF$.
   \label{fig:difphasecom}}
   \end{figure}

There have been suggestions that the effect of the glitch associated
with the second helium ionization zone varies with the phase of the solar cycle 
\citep{Gough1995, Gough2002b, Basu2004, Ballot2006, Verner2006}.
We would expect that such a variation should be visible also in $\CF$.
To investigate this, Fig.~\ref{fig:difphasecom} combines the
fitted $\CH_2$ for all 72-d observations, after subtraction of the linear
fit.
There is evidently very little scatter in the result, and further
inspection shows no evidence for systematic variations with the
phase of the solar cycle.
Similarly, the second differences in the smooth component of the frequencies
(see the lower panel of Fig. \ref{fig:signal_sun}) show no significant
variation with solar cycle.
To set the scale, we note that the change $\delta \nu$ in
cyclic frequency, corresponding to a change $\delta \CF$ in $\CF$,
approximately satisfies
\begin{equation}
\delta \nu \simeq - {\Delta \nu \over \pi} \delta \CF \; .
\end{equation}
{}From this we estimate that the range of the residual variation in
$\nu$ with solar cycle, after removing the linear trend, is below
$0.04 \muHz$, which apparently is substantially smaller than
the variation in the signature of helium ionization inferred
by \citet{Basu2004} and \citet{Verner2006}.
This deserves further investigation.

\label{lastpage}
\end{document}